\documentclass[twocolumn]{aastex62}

\usepackage{orcidlink}
\usepackage{amsmath}
\usepackage{hyperref}
\usepackage{xcolor}

\newcommand{\comcorrect}[1]{\textcolor{black}{#1}}
\newcommand{\cwr}[1]{\textcolor{black}{ #1}}
\begin{document}
\newcommand{\refco}[1]{\textcolor{black}{#1}}

\date{Sep 2025}
\title{The Draco Dwarf Spheroidal Galaxy in the First Year of DESI Data}

\author[0000-0003-4651-8510]{J.~Ding}
\affiliation{Department of Astronomy and Astrophysics, University of California, Santa Cruz, 1156 High Street, Santa Cruz, CA 95064, USA}
\affiliation{Steward Observatory, University of Arizona, 933 N Cherry Avenue, Tucson, AZ 85721, USA}
\author[0000-0002-6667-7028
]{C.~Rockosi}
\affil{Department of Astronomy and Astrophysics, University of California, Santa Cruz, 1156 High Street, Santa Cruz, CA 95064, USA}
\author[0000-0002-9110-6163]{Ting~S.~Li}
\affil{Department of Astronomy \& Astrophysics, University of Toronto, Toronto, ON M5S 3H4, Canada}
\author[0000-0003-2644-135X]{S.~E.~Koposov}
\affil{Institute for Astronomy, University of Edinburgh, Royal Observatory, Blackford Hill, Edinburgh EH9 3HJ, UK}
\affil{Institute of Astronomy, University of Cambridge, Madingley Rd, Cambridge CB3 0HA, UK}
\author[0000-0001-5805-5766
]{A.~H.~Riley}
\affil{Institute for Computational Cosmology, Department of Physics, Durham University, South Road, Durham DH1 3LE, UK}
\author{W.~Wang}
\affil{Department of Astronomy, School of Physics and Astronomy, Shanghai Jiao Tong University, Shanghai 200240, China}
\author[0000-0001-8274-158X
]{A.~P.~Cooper}
\affil{Institute of Astronomy and Department of Physics, National Tsing Hua University, 101 Kuang-Fu Rd. Sec. 2, Hsinchu 30013, Taiwan}
\affil{Center for Informatics and Computation in Astronomy, National Tsing Hua University, Taiwan}
\author{N.~Kizhuprakkat}
\affil{Institute of Astronomy and Physics Department, National Tsing Hua University, 101 Kuang Fu Rd. Sec. 2, Hsinchu 30013, Taiwan}
\affil{Center for Informatics and Computation in Astronomy, National Tsing Hua University, Taiwan}
\author[0000-0002-2527-8899
]{M.~Lambert}
\affil{Department of Astronomy and Astrophysics, University of California, Santa Cruz, 1156 High Street, Santa Cruz, CA 95064, USA}
\author[0000-0003-0105-9576]{G.~E.~Medina}
\affil{Department of Astronomy and Astrophysics, University of Toronto, 50 St. George Street, Toronto, ON M5P 0A2, Canada}
\author[0000-0002-7393-3595
]{N.~Sandford}
\affil{Department of Astronomy and Astrophysics, University of Toronto, 50 St. George Street, Toronto, ON M5P 0A2, Canada}
\author[0000-0003-0822-452X]{J.~Aguilar}
\affil{Lawrence Berkeley National Laboratory, 1 Cyclotron Road, Berkeley, CA 94720, USA}
\author[0000-0001-6098-7247]{S.~Ahlen}
\affil{Physics Dept., Boston University, 590 Commonwealth Avenue, Boston, MA 02215, USA}
\author[0000-0001-9712-0006]{D.~Bianchi}
\affil{Dipartimento di Fisica ``Aldo Pontremoli'', Universit\`a degli Studi di Milano, Via Celoria 16, I-20133 Milano, Italy}
\affil{INAF-Osservatorio Astronomico di Brera, Via Brera 28, 20122 Milano, Italy}
\author[0000-0002-8458-5047]{D.~Brooks}
\affil{Department of Physics \& Astronomy, University College London, Gower Street, London, WC1E 6BT, UK}
\author{T.~Claybaugh}
\affil{Lawrence Berkeley National Laboratory, 1 Cyclotron Road, Berkeley, CA 94720, USA}
\author[0000-0002-1769-1640]{A.~de la Macorra}
\affil{Instituto de F\'{\i}sica, Universidad Nacional Aut\'{o}noma de M\'{e}xico, Circuito de la Investigaci\'{o}n Cient\'{\i}fica, Ciudad Universitaria, Cd. de M\'{e}xico C.~P.~04510, M\'{e}xico}
\author{P.~Doel}
\affil{Department of Physics \& Astronomy, University College London, Gower Street, London, WC1E 6BT, UK}
\author[0000-0002-2890-3725]{J.~E.~Forero-Romero}
\affil{Observatorio Astron\'omico, Universidad de los Andes, Cra. 1 No. 18A-10, Edificio H, CP 111711 Bogot\'a, Colombia}
\author[0000-0001-9632-0815]{E.~Gaztañaga}
\affil{Institute of Cosmology and Gravitation, University of Portsmouth, Dennis Sciama Building, Portsmouth, PO1 3FX, UK}
\author[0000-0003-3142-233X]{S.~Gontcho A Gontcho}
\affil{Lawrence Berkeley National Laboratory, 1 Cyclotron Road, Berkeley, CA 94720, USA}
\author{G.~Gutierrez}
\affil{Fermi National Accelerator Laboratory, PO Box 500, Batavia, IL 60510, USA}
\author[0000-0001-9822-6793]{J.~Guy}
\affil{Lawrence Berkeley National Laboratory, 1 Cyclotron Road, Berkeley, CA 94720, USA}
\author[0000-0002-6024-466X
]{M.~Ishak}
\affil{Department of Physics, The University of Texas at Dallas, 800 W. Campbell Rd., Richardson, TX 75080, USA}
\author[0000-0002-7101-697X]{R.~Kehoe}
\affil{Department of Physics, Southern Methodist University, 3215 Daniel Avenue, Dallas, TX 75275, USA}
\author[0000-0003-3510-7134]{T.~Kisner}
\affil{Lawrence Berkeley National Laboratory, 1 Cyclotron Road, Berkeley, CA 94720, USA}
\author[0000-0001-6356-7424]{A.~Kremin}
\affil{Lawrence Berkeley National Laboratory, 1 Cyclotron Road, Berkeley, CA 94720, USA}
\author{O.~Lahav}
\affil{Department of Physics \& Astronomy, University College London, Gower Street, London, WC1E 6BT, UK}
\author[0000-0003-1838-8528
]{M.~Landriau}
\affil{Lawrence Berkeley National Laboratory, 1 Cyclotron Road, Berkeley, CA 94720, USA}
\author[0000-0001-7178-8868]{L.~Le~Guillou}
\affil{Sorbonne Universit\'{e}, CNRS/IN2P3, Laboratoire de Physique Nucl\'{e}aire et de Hautes Energies (LPNHE), FR-75005 Paris, France}
\author[0000-0002-1125-7384]{A.~Meisner}
\affil{NSF NOIRLab, 950 N. Cherry Ave., Tucson, AZ 85719, USA}
\author[0000-0002-6610-4836]{R.~Miquel}
\affil{Instituci\'{o} Catalana de Recerca i Estudis Avan\c{c}ats, Passeig de Llu\'{\i}s Companys, 23, 08010 Barcelona, Spain}
\author[0000-0002-2733-4559]{J.~Moustakas}
\affil{Department of Physics and Astronomy, Siena College, 515 Loudon Road, Loudonville, NY 12211, USA}
\author[0000-0001-7145-8674]{F.~Prada}
\affil{Instituto de Astrof\'{i}sica de Andaluc\'{i}a (CSIC), Glorieta de la Astronom\'{i}a, s/n, E-18008 Granada, Spain}
\author[0000-0001-6979-0125]{I.~P\'erez-R\`afols}
\affil{Departament de F\'isica, EEBE, Universitat Polit\`ecnica de Catalunya, c/Eduard Maristany 10, 08930 Barcelona, Spain}
\author{G.~Rossi}
\affil{Department of Physics and Astronomy, Sejong University, 209 Neungdong-ro, Gwangjin-gu, Seoul 05006, Republic of Korea}
\author[0000-0002-9646-8198]{E.~Sanchez}
\affil{CIEMAT, Avenida Complutense 40, E-28040 Madrid, Spain}
\author{M.~Schubnell}
\affil{University of Michigan, 500 S. State Street, Ann Arbor, MI 48109, USA}
\author[0000-0002-3461-0320]{J.~Silber}
\affil{Lawrence Berkeley National Laboratory, 1 Cyclotron Road, Berkeley, CA 94720, USA}
\author{D.~Sprayberry}
\affil{NSF NOIRLab, 950 N. Cherry Ave., Tucson, AZ 85719, USA}
\author[0000-0003-1704-0781]{G.~Tarl\'{e}}
\affil{University of Michigan, 500 S. State Street, Ann Arbor, MI 48109, USA}
\author{B.~A.~Weaver}
\affil{NSF NOIRLab, 950 N. Cherry Ave., Tucson, AZ 85719, USA}
\author[0000-0001-5381-4372]{R.~Zhou}
\affil{Lawrence Berkeley National Laboratory, 1 Cyclotron Road, Berkeley, CA 94720, USA}

\begin{abstract}
We investigate the spatial distribution, kinematics, and metallicity of stars in the Draco dwarf spheroidal galaxy using data from the Dark Energy Spectroscopic Instrument (DESI).  We identify 155 high probability members of Draco using line of sight velocity and metallicity information derived from DESI spectroscopy along with {\it Gaia} DR3 proper motions. We find a mean line of sight velocity of $ -290.62\pm0.80$ km s$^{-1}$ with dispersion = $9.57^{+0.66}_{-0.62}$ km s$^{-1}$ and mean metallicity $\rm{[Fe/H]}$ = $-2.10\pm0.04$, consistent with previous results.  We also find that Draco has a steep metallicity gradient within the half-light radius, and a metallicity gradient that flattens beyond the half-light radius. We identify eight high probability members outside the King tidal radius, \cwr{four of which we identify for the first time.} These extra-tidal stars are not preferentially aligned along the orbit of Draco. We compute an average surface brightness of \refco{34.02} mag $\rm arcsec^{-2}$ within an elliptical annulus from the King tidal radius of 48.1 arcmin to 81 arcmin. 
\end{abstract}
\section{Introduction} \label{intro}
Dwarf galaxies play a crucial role in our understanding of galaxy formation and evolution. They are key to understanding the dark matter distribution, star formation history, chemical and dynamical evolution of the Milky Way (MW) 
\citep[e.g.,][]{Benson2002,Bullock2005,Kirby2008,Fattahi2020}. Starting with the classical dwarf spheroidal galaxies (dSphs) of the MW, the discovery of Sagittarius dSph \citep{Ibata1994} was followed by many new, faint dwarf galaxies discovered at a rapid pace in the Sloan Digital Sky Survey \comcorrect{(SDSS)} \citep[e.g.,][]{Willman2005,Belo2006,Belo2007,Belo2008,Koposov2008}.
The recent discovery of new \refco{dSphs} in the Dark Energy Survey \comcorrect{(DES)} \citep{Bechtol2015,Koposov2015des,Wagner2015,McNanna2024}, Magellanic Satellites Survey \citep{Torrealba2018} and other surveys \citep[e.g.,][]{Drlica-Wagner2021,Wright2024} has revolutionized our knowledge of the MW's satellite systems. These findings reveal a large population of dSphs, with some being potential satellites of the Large Magellanic Cloud (LMC) and the Small Magellanic Cloud (SMC) \citep{Pardy2020,Kalli2018}.

\refco{dSphs} live in low-mass dark matter halos and are the most dark matter dominated systems in the Universe today \citep[e.g.,][]{Pryor1990}.  The subhalo mass function of a galaxy like the MW and the central density profiles of \refco{dSphs} and their subhalos are among the few measurements that can constrain the distribution of dark matter on small scales \citep{Tau2025,Cooper2025}. Thus dwarf galaxies play an important role in investigating the nature of dark matter itself \citep{Ostriker2003,Zavala2019}.  Data from the {\it Gaia} satellite \citep{Gaia2018} and recent photometric surveys have enabled new investigations of dwarf galaxy kinematics and their interactions with the MW \citep[][]{Bechtol2015,Mu20182,Li2018,Shipp2018,Pace&Li2019,Pace2022,Hammer2023}. These and other recent investigations suggest that some of these dwarfs may be surrounded by significant low-surface-brightness stellar tidal features, a phenomenon predicted by simulations even for galaxies not currently experiencing significant mass loss through tidal stripping \citep{cooper2010,Wang2017,Riley2024, Shipp2024}. Many studies have found evidence of stars being stripped from MW dSph satellite galaxies and potential associations of dSph galaxies with extended stellar streams \citep[e.g.,][]{New2002,Maj2003,Law2005,Kop2012,Ibata2020,Kado-Fong2020,Vas2021,Vivas2022}.  

Another possible explanation for the extended stellar structures around \refco{dSphs} may be that they are stellar halos. Simulations show that dwarf galaxies experience mergers that leave behind remnant stellar halos \citep{Tarumi2021,Deason2022,Goater2024}. Several recent studies have found spatially extended low surface brightness structure around dwarf galaxies \citep{Chiti2021,Yang2022,Jensen2024,Tau2024,Conroy2024} and attribute them to stellar halos.

Among dwarf galaxies, Draco was first discovered by \cite{Wilson1955} using the Palomar 48 inch telescope. Draco is a metal poor dwarf, [Fe/H] = $-1.93\pm0.01$ dex as measured by \cite{Kirby2011}.  It is at a heliocentric distance of 76 kpc, with a half-light radius of 9.67 arcmin and a King tidal radius of 48.1 arcmin \citep{Mu20182}.  It is highly dark matter dominated \citep{Kleyna2002,Klessen2003,Vitral2024,Yang2025}. Its large velocity dispersion has made Draco the subject of interest in many previous studies, none of which found clear evidence for tidal disruption in Draco \citep{Segall2007,Mu20182,Jensen2024}. \cwr{\citet{Qi2022} and \citet{Jensen2024} investigated the density profile and other properties of Draco using Gaia data.  \cite{Qi2022} find nine candidate Draco members beyond its tidal radius. \cite{Jensen2024} also use Gaia data to search for but do not find evidence for an extended, extra-tidal spatial component around Draco.  The availability of spectroscopic data is an an opportunity to further investigate the properties of the Draco dSph, particularly in the outskirt region.}

\comcorrect{The primary goal for the Dark Energy Spectroscopic Instrument (DESI)} is the precise measurement of dark energy and the expansion history of the Universe, and this is the focus of observations under dark sky conditions with good transparency \citep[e.g.,][]{levi2013,DESI2016I,DESI2022int,Miller2024,Poppett2024}. DESI also executes a bright time survey when the moon is bright and in conditions of moderate seeing and/or transparency \citep{Schlafly2023}. This bright survey is dedicated to studying the MW and brighter galaxies for the primary cosmology science program \citep{DESI2016}.
Before the start of the main bright and dark time surveys described above, DESI engaged in a Survey Validation (SV) campaign that included the MW survey. The SV campaign \citep{DESI2023b} observed $>$ 200,000 unique stars. Data from the SV campaign is included in the DESI Early Data Release (EDR) \citep{DESI2023a}.
 The Draco dwarf was observed on two DESI tiles during the SV campaign.
 The DESI bright time survey tiles taken in the first year of the survey also cover an area that overlaps with these two Draco SV tiles. 
 
This paper uses the year-1 and SV DESI data.  These data were included in the DR1 DESI public data release \citep{DESIDR12025}. We note the majority of the Draco observations were taken from the two tiles observed during SV.
 With the addition of heliocentric velocity and metallicity measurements \citep{Cooper2023,DESI2023a} from DESI in a large area around Draco, we revisit the properties of the Draco dSph especially beyond the King tidal radius. \cwr{We adopt a King tidal radius of 48.1 arcmin as measured by \cite{Mu20182} in this paper.} 

 
This paper is organized as follows. In \S{\ref{data}}, we explain our sample selection. In \S{\ref{method}}, we discuss the Gaussian mixture model (GMM) used to identify member stars in Draco. In \S{\ref{results}}, we illustrate the membership probabilities and the line of sight velocity,  metallicity, and proper motion properties of Draco from that analysis.  In \S{\ref{discussion}} and \S{\ref{conclusion}}, we compare our results to the previous literature and present our conclusions. 

\section{Data \label{data}}
The Draco data used in this study were taken as part of the DESI Science Verification program \citep{DESI2023b} and the first year of the DESI Milky Way Survey. The SV spectra are included in the DESI Early Data Release (EDR) \citep{DESI2023a}.  The DESI Milky Way  Survey data were included in DR1 \citep{DESIDR12025}. \cwr{The selection of MW Survey spectroscopic targets for the DESI Milky Way Survey is described in \citet{Cooper2023}. In summary, the highest priority targets are $16<r<19$ blue halo tracers and redder stars selected using an additional loose Gaia parallax cut to prioritize red giants in the stellar halo.}

\cwr{Radial velocities and metallicities for the SV and year-1 data are included in the DESI DR1 Stellar Catalog \citep{Koposov2025} and use an improved version of the \textsc{rvspecfit} pipeline compared to the one discussed in \citet{Koposov2024}. The catalog includes the fiber map and cross matched {\it Gaia} information as discussed in \citet{Cooper2023} and \citet{Koposov2024}.}

The two DESI SV tiles cover the same 8 square degree DESI field of view.  The spectroscopic targets for the Draco SV tiles were selected from {\it Gaia} DR3 \citep{Gaia2023} and \comcorrect{the Dark Energy Camera Legacy Survey (DECaLS)} DR9 \citep{Dey2019} over the entire tile area.  The color and magnitude limits are 16 $< r <$ 21 and $g-r$ $<$ 1.2. We select Draco candidate members for DESI SV spectroscopic observation as follows:

{\noindent 1. $\rm |{\bf{pm}}-{\bf{pm0}}| < 2 \text{ mas yr}^{-1}$, where ${\bf{pm}}$ is the stellar proper motion vector and ${\bf{pm0}}$ is the Draco dSph proper motion: $\rm \mu_{\alpha} cos\delta = 0.044$ mas ${\rm yr^{-1}}$ and $\rm \mu_{\delta} = -0.188$ mas ${\rm yr^{-1}}$ from \cite{Pace2022}. }

{\noindent 2. $\omega$ $-$ 3 $\times \rm \sigma_{\omega} < 1/D$, where $\omega$ is the parallax, $\rm \sigma_{\omega}$ is the parallax error and D is the distance to Draco of 75.8 kpc \citep{Mu20182}.}

{\noindent 3. Star/galaxy separation cut, as discussed in \citet{Riello2021} and \citet{Pace2022}: }
\begin{itemize}
 \item \tt{gaia$\_$astrometric$\_$excess$\_$noise} $<$ 1 
 \item \tt{gaia\_phot\_bp\_rp\_excess\_factor} $<1.3$ + \\
         $0.06\times$(\tt{gaia\_phot\_bp\_mean\_mag} 
        $-$  \\ \tt{gaia\_phot\_rp\_mean\_mag})$^{2}$.
 \end{itemize}
\cwr{These terms are defined in the $Gaia$ data release documentation \footnote{https://gea.esac.esa.int/archive/documentation/GDR2/}.}

 \cwr{Each tile targets 5000 objects}, including calibration and sky fibers. See Section 2.2 in \citet{DESI2023a} for more details of the DESI observation.  The two Draco SV tiles were observed in a range of moon illumination and observing conditions. The median actual exposure time for the stars used for this analysis was 6100 s. For the main survey Draco stars used in the analysis the median effective exposure time is 265 s.  For the SV Draco data used here, the median effective exposure time\footnote{\cwr{The spectroscopic effective exposure time, which is corrected to nominal DESI observing conditions, is defined by equation 22 in \cite{Guy2023}.}} is 7760 s.  

From the spectroscopic sample, we make the following additional selection of Draco candidates in RA/DEC, data quality, and stellar parameters to limit the number of non-Draco stars in our sample. Note that many of the DESI data quality cuts are as advised in \cite{Koposov2024}. 
\begin{itemize}
\item RA from 253 to 267 deg
\item Dec from 55.9 to 61.0 deg
\item 16 $<$ $r$ $<$ 21
\item $\tt{log \it{g}}$ $<$ 4
\item ${\tt RVS\_WARN}$ = 0
\item $\tt{RR\_SPECTYPE}$ not QSO
\item $\tt{PHOT\_VARIABLE\_FLAG}$ not VARIABLE
\item \cwr{has a Gaia proper motion measurement}
\end{itemize}
\cwr{Duplicates in the sample are removed using the unique {\it Gaia} source ID.} The total number of stars after this preliminary selection is 736. 
\subsection{Color-Magnitude Selection \label{2.1}}

\begin{figure*}
\centering
\includegraphics[width=12cm]{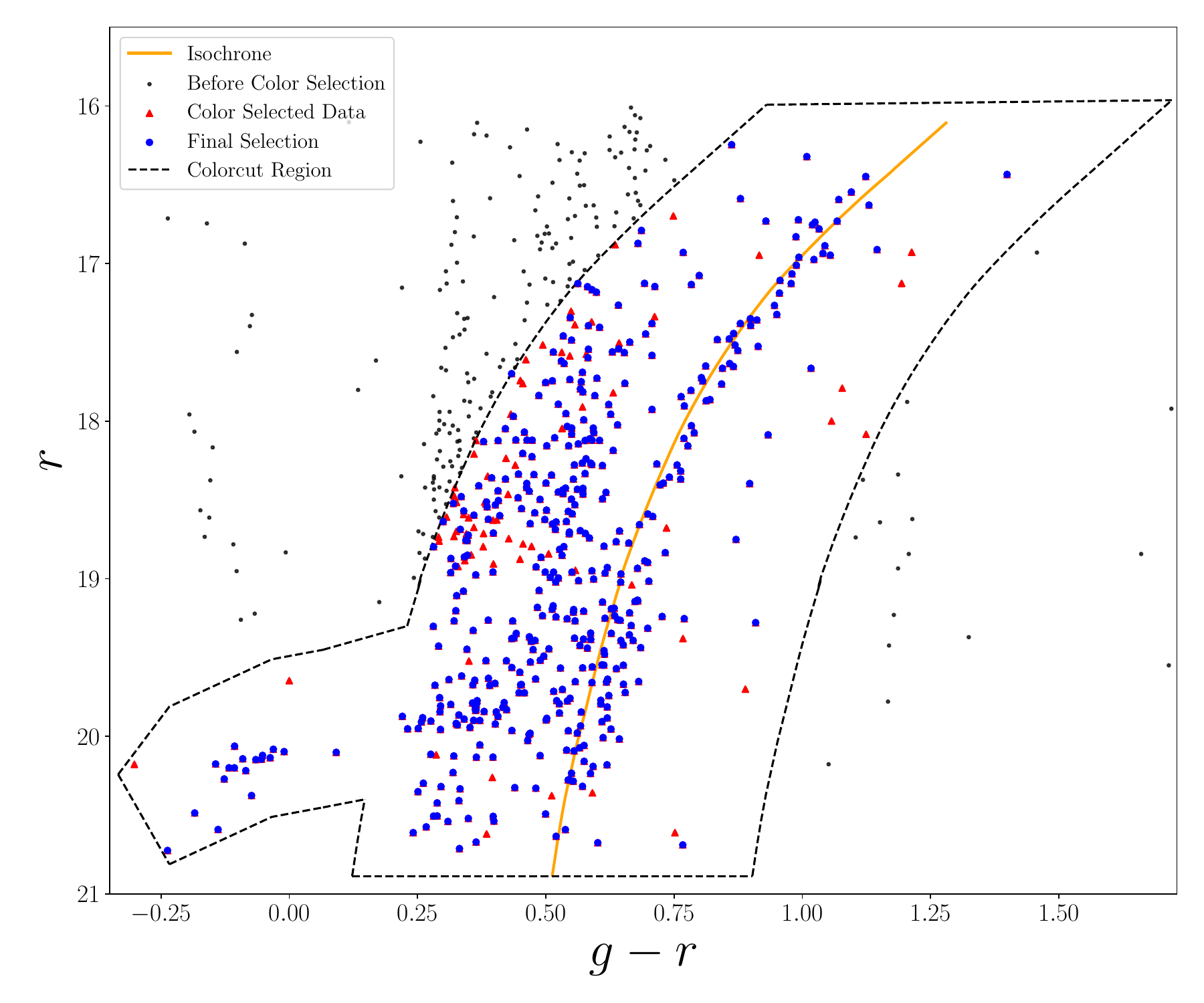}
\caption{\comcorrect{Color-magnitude diagram of our DESI Draco sample.} The region inside the dashed lines is the color selection for the Draco candidates sample from the DESI DR1 Stellar Catalog data. For the color-magnitude cuts used in the selection, we use an old, metal-poor ([Fe/H] = $-$1.5 dex) Dartmouth isochrone with an age = 10 Gyr \citep{Dotter2008} and select stars within $\pm$0.39 mag of the Isochrone $g-r$ value. We use Legacy survey \citep{Dey2019} photometry of the M92 globular cluster to define a region for the horizontal branch, since that is not included in the isochrone. The data sample after the color cut is labeled in red triangle, and the final selected data sample (see \S{\ref{2.1}}) is labeled in blue circle. \label{fig:colorcut}}
\end{figure*}

We use photometric information from the Legacy Survey DR9 \citep{Dey2019} to improve our selection of Draco candidate members using color-magnitude cuts. We use an old (age = 10 Gyr), metal-poor ([Fe/H] = $-1.5$) Dartmouth Isochrone \citep{Dotter2008}. \refco{We selected this isochrone because it is the best fit to the Draco members within one half-light radius. Candidate stars are selected within $|g-r-Iso(r)|<$ 0.39, where $Iso(r)$ is the isochrone $g-r$ value at any value of $r$. We adopt the value of 0.39 because it removes the most background stars from the color-magnitude diagram (CMD) while including the Draco members.} The horizontal branch is not included in the Dartmouth Isochrone model. We use Legacy Survey \citep{Dey2019} photometry of the globular cluster M92 to define a region for the horizontal branch of an old stellar population like Draco.  We select a region 0.6 mag wide in the $r$ band centered on the cluster horizontal branch ridgeline and add that to the isochrone selection. \comcorrect{All of the photometry has been extinction-corrected.} \cwr{The CMD selection is illustrated in Fig. \ref{fig:colorcut}.} After CMD selection, the total number of stars remaining in our candidate sample is 522. We then apply a velocity cut from $-500$ to 100 km $\rm s^{-1}$ (the Draco mean heliocentric velocity is -291 km $\rm s^{-1}$ from \cite{Mu20182}), and a metallicity cut $-4.0$ $<$ [Fe/H] $<$ $-$0.2 dex. \refco{The CMD selection removes almost all the Milky Way interlopers. Using this metallicity range combined with other selection, we can model the background star distribution well with a single Gaussian.} We also apply a proper motion cut between $-3.5$ and 3 mas $\rm yr^{-1}$ since this cut was not applied when selecting DR1 targets. \refco{214 stars were excluded by the color cut from the preliminary selection, six by the metallicity cut, 63 by the proper motion cut, and three by the velocity cut.} The final selection yields 450 stars, 100 from the first year of the DESI main survey, and the rest from the SV data. The final sample is shown in Fig. \ref{fig:colorcut} as blue circles.

\section{Methods \label{method}}
We apply a GMM to the combined dataset of \comcorrect{heliocentric radial velocity} $v\rm{_{hel}}$ and [Fe/H] from DESI and proper motions from {\it Gaia} DR3. We correct the radial velocity for perspective rotation following \cite{Kap2008}. We implement the correction as part of fitting the mixture model parameters, so the correction is made using our best-fit proper motion.  All the radial velocity and proper motion data are in the heliocentric frame. We model the radial velocity,  $\rm{[Fe/H]}$ and the proper motion distributions as a two-component mixture of the satellite and MW foreground and background interlopers. We do not include a radial density profile in our mixture model in order to avoid  biasing our membership probabilities against stars at large radius.
 Our model has two components, Draco and MW interlopers, and uses a Gaussian function to describe three parameters: $v\rm{_{hel}}$, [Fe/H], and proper motion, for each component.  The functional form for the likelihood of each component is:
\begin{gather}
L(\overline{x}, \sigma^2; x) = \frac{1}{\sqrt{2\pi\sigma^2}} \exp\left(-\frac{(x - \overline{x})^2}{2\sigma^2}\right),
\end{gather}
 where $\overline{x}$ is the mean of the physical property and $\sigma$ is the combination of the intrinsic dispersion of the physical property and the internal errors of the data. 
 The two-component mixture model is a sum of likelihoods:

\begin{gather}
L = (f_{\text{satellite}}) L_{\text{satellite}} + (1-f_{\text{satellite}}) L_{\text{MW}},\\
L_{\text{satellite}} = L_{\text{$v\rm{_{hel}}$,satellite}}L_{\text{$\rm{[Fe/H]}$,satellite}}L_{\text{PM,satellite}},\\
L_{\text{MW}} = L_{\text{$v\rm{_{hel}}$,MW}} L_{\text{$\rm{[Fe/H]}$},\rm{MW}}L_{\text{PM,MW}},
\end{gather}
where $f_{\text{satellite}}$ is the fraction of stars belonging to the dSph. Variables $L_{\text{$v\rm{_{hel}}$,satellite}}$, $L_{\text{$\rm{[Fe/H]}$,satellite}}$, $L_{\text{PM,satellite}}$ refer to the heliocentric velocity, metallicity and proper motion likelihood components for the dSph. $L_{\text{$v\rm{_{hel}}$,MW}}$, $L_{\text{$\rm{[Fe/H]}$,MW}}$, $L_{\text{PM,MW}}$ denote the heliocentric velocity, metallicity and proper motion likelihood components for the MW interloper stars. \cwr{ Note that the components of $L_{\text{satellite}}$ and  $L_{\text{MW}}$ are the products of the Gaussian likelihood function in equation (1) evaluated for each individual star. }

We list all the parameters used in the model in Table \ref{tab:prior}. The four parameters ($v\rm{_{hel,satellite}}$, $\rm{[Fe/H]_{satellite}}$, $\mu_{\rm{\alpha}}\cos\delta,\rm_{satellite}$, 
  $\mu_{\rm{\delta,satellite}}$) represent the mean of the physical properties for the dSph component and $v\rm{_{hel,MW}}$, $\rm{[Fe/H]_{MW}}$, $\mu_{\rm \alpha} \cos\delta,_{\rm\rm{MW}}$,   
    ${\mu_{\delta,\rm{MW}}}$
represent the mean for the MW interlopers. We fix the dispersion of Draco's proper motion ($\sigma_{\mu_{\alpha }\cos_\delta\rm, satellite}$ and $\sigma_{\mu_{\delta},\rm satellite}$) using the measurement of Draco's line of sight velocity dispersion by \cite{Mu20182}, \cwr{because the proper motion errors are much larger than the projected internal velocity dispersion of Draco.} We convert their measurement of the Draco velocity dispersion of 9.1 km s$^{-1}$ to proper motion using their distance to Draco of 75.8 kpc. We then fix the dispersion of Draco's proper motion in our model to the resulting value of 0.025 mas yr$^{-1}$.  Therefore, we have two parameters $\sigma_{ v\rm_{hel,satellite}}$, ${{\sigma_{\rm [Fe/H],satellite}}}$ for the dSph dispersion and four parameters, $\sigma_{v\rm_{hel,MW}}$, $\rm{\sigma_{[Fe/H],MW}}$, $\rm{\sigma_{{\mu_{\alpha}\cos \delta,_{\rm MW}}}}$ and $\rm{\sigma_{\mu_{\delta,MW}}}$ for the MW dispersion. Also, as discussed in section 7.4.2 in \cite{Cooper2023}, there is an additional systematic component of $\sim$ 0.9 km s$^{-1}$ in the DESI RVS radial velocity errors.  Therefore, we add 0.9 km s$^{-1}$ in quadrature with the radial velocity errors in our GMM. 

\begin{figure}
\centering
\includegraphics[width=8.8cm]{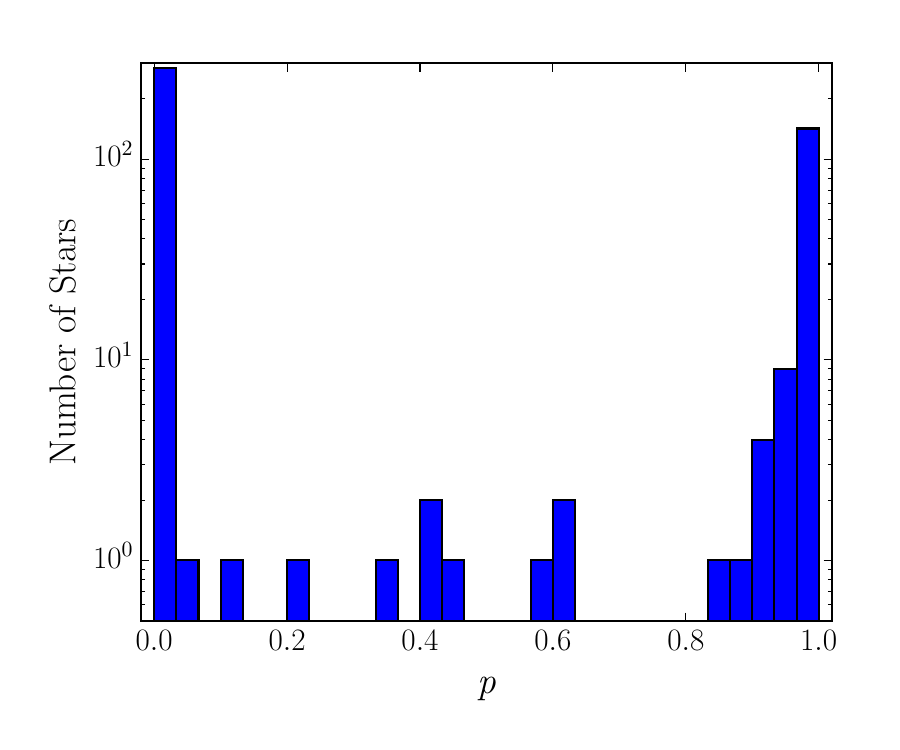}
\caption{The probability distribution for Draco members identified with the GMM. From this probability distribution, we define the high probability members to be $p$ $>$ 0.9. There are \comcorrect{155} stars that pass this criterion. \label{fig:probdis}}
\end{figure}

We model the proper motion of the Draco and MW components as a multivariate distribution in Right Ascension and Declination \citep[e.g.,][]{Pace2022}. We include the covariance terms ($\text{Cov}(\mu_{\delta},\mu_{\alpha}\cos \delta )$) in the proper motion errors 
and internal proper motion dispersion for each component. 
The log likelihood for this multivariate Gaussian distribution is:

\begin{equation}
\begin{split}
L(\boldsymbol{\overline{x}}, \boldsymbol{M} \mid \mathbf{X}) =
-\frac{n}{2} \log(2\pi) -\frac{1}{2} \log|\boldsymbol{M} | 
\\
- 
\frac{1}{2} \sum_{i=1}^{n} (\mathbf{x}_i- \boldsymbol{\overline{x}})^T \boldsymbol{M}^{-1} (\mathbf{x}_i - \boldsymbol{\overline{x}})
\end{split}
\end{equation}
where $\boldsymbol{\overline{x}}$ represents the parameters for the mean proper motion ($\mu_{\delta}$, $\mu_{\alpha} \cos \delta$), \comcorrect{$\rm \bf{M}$} represents the covariance matrix, $\mathbf{X}$ represents the observed proper motion data, and $n$ is the number of stars in the sample. 
For the proper motion distribution, we include the intrinsic dispersion by adding the dispersion component to the diagonal term of the PM covariance. 
The covariance matrix ${\boldsymbol{\rm M}}$ is defined as: 
\begin{equation}
\text{Cov}(\mu_{\delta},\mu_{\alpha}\cos \delta ) + \rm Diag(\sigma_{(\mu_{\alpha}\cos\delta)}, \sigma_{\mu_\delta})
\end{equation}


\noindent where $\text{Cov}(\mu_{\delta}, \mu_{\alpha}\cos \delta)$, is the {\it Gaia} covariance matrix for proper motions. For the Draco component, we add $\sigma_{\mu_{\alpha}\cos \delta,{\mathrm{satellite}}}$ and $\mathrm{\sigma_{\mu_{\delta},satellite}}$,  the mean Draco proper motion dispersion we fixed as explained above, on the diagonal. For the MW interlopers component, we add ${\sigma_{\mu_{\alpha}\cos\delta,{_\mathrm{MW}}}}$ and $\mathrm{\sigma_{\mu_{\delta,MW}}}$, the parameters fit by the mixture model, on the diagonal.  

From the above discussion, there are 15 parameters in total for the model (see Table \ref{tab:prior}). In order to estimate the best values for the parameters, we run an MCMC process based on the affine invariant ensemble sampler for Markov Chain Monte Carlo \citep{Goodman2010} in the $\it{emcee}$ library \citep{Mackey2013}. The prior distributions for all the parameters are as follows (see Table \ref{tab:prior}):
 \newline
(i) {\textbf{$f_{\text{sat}}$ }}: One uniform prior between 0 and 1 for the fraction $f_{\text{sat}}$ defined above. \\
\cwr{(ii) $\boldsymbol{v_{\mathrm{hel,satellite}}}, \
\boldsymbol{v_{\mathrm{hel,MW}}}, \ 
\sigma_{\boldsymbol{v_{\mathrm{hel,satellite}}}}, \
\sigma_{\boldsymbol{v_{\mathrm{hel,MW}}}}$: Four parameters describing the mean and dispersion of the heliocentric velocity, two for the velocity distribution of the Draco component and two for the MW component. We use a uniform distribution prior for both the Draco component and the MW interlopers with a mean heliocentric velocity between $-$500 and 100 km $\rm s^{-1}$ and log dispersion between $-$1 and 3. \\
(iii) {$\rm{\bf [Fe/H],satellite}$, $\bf \rm{\bf [Fe/H],MW}$, ${\rm{\bf \sigma_{[Fe/H],satellite}}}$, $\bf \rm{\bf \sigma_{[Fe/H],MW}}$
}: Four parameters describing the mean and dispersion of $\rm{[Fe/H]}$. For both the Draco component and the MW interlopers component, we use a uniform distribution prior for the mean $\rm{[Fe/H]}$ between $-$4.0 and $-$0.2 dex and log dispersion between $-$3 and 1. \\
(iv) $\boldsymbol{\mu_{\alpha} \cos \delta_{\mathrm{,\ satellite}}}, \
\boldsymbol{\mu_{\delta,\mathrm{satellite}}}$: Two parameters describing the Draco Proper motion. We use a uniform distribution prior for the mean proper motion for both $\mu_{\alpha} \cos \delta,_{\rm satellite}$ and $\mu_{\delta,\rm satellite}$ between $-$3.5 and 3 mas $\rm yr^{-1}$. Note that the dispersion of $\mu_{\alpha} \cos \delta,_{\rm satellite}$ and $\mu_{\delta,\rm satellite}$ for the Draco component are fixed to 0.025 mas $\rm yr^{-1}$, as discussed above.\\
(v) $\boldsymbol{\mu_{\alpha} \cos \delta_{\mathrm{,\ MW}}}, \
\boldsymbol{\mu_{\delta,\mathrm{MW}}}, \
\sigma_{\boldsymbol{\mu_{\alpha} \cos \delta_{\mathrm{,\ MW}}}}, \
\sigma_{\boldsymbol{\mu_{\delta,\mathrm{MW}}}}$: Four parameters describing the MW Proper motion. We use a uniform distribution prior for the mean proper motion $\mu_{\alpha} \cos \delta,_{\rm MW}$ and $\mu_{\delta,\rm MW}$ between $-$3.5 and 3 mas $\rm yr^{-1}$ and log dispersion for $\mu_{\alpha} \cos \delta,_{\rm MW}$ and $\mu_{\delta,\rm MW}$ between $-$1 and 1.3. }

These priors are summarized in Table \ref{tab:prior}.
With the $v\rm{_{hel}}$, [Fe/H] and proper motion likelihood we described above, for each star, we can compute the relative likelihood of membership in Draco and the MW \refco{based on equation (1)-(4).} This gives us membership probabilities for the stars in Draco. The probability that a star belongs to the satellite is:
\begin{equation}
p = \frac{{(f_{\text{sat}}) L_{\text{satellite}}}}{(f_{\text{sat}}) L_{\text{satellite}} + (1-f_{\text{sat}}) L_{\text{MW}}}
\end{equation}
where the component $L_{\text{satellite}}$ refers to the dSph and the component $L_{\text{MW}}$ describes the MW interlopers stars. 


\begin{deluxetable*}{ccc}[!t]
\tablecaption{Priors and Best Fit Parameters for the GMM\label{tab:prior}}
\tablewidth{0pt}
\tablehead{
\colhead{Parameters} &
\colhead{Prior Range} &
\colhead{Best fit value with error}
}
\startdata
$f_{\text{sat}}$ & (0,1) &$0.36\pm0.02$\\
$v\rm{_{hel,satellite}}$ (km s$^{-1}$) & ($-$500,100) & $-290.62\pm0.80$ \\
$v\rm{_{hel,MW}}$ (km s$^{-1}$) & ($-$500,100) & $-185.37\pm6.19$\\
$\sigma_{ v\rm_{hel,satellite}}$ (km s$^{-1}$) & (0.1,1000) & $9.57^{+0.66}_{-0.62}$ \\
$\sigma_{v\rm_{hel,MW}}$ (km s$^{-1}$)  & (0.1,1000) & $107.15^{4.53}_{4.35}$ \\
$\rm{[Fe/H],satellite}$
& ($-$4.0, $-$0.2)& $-2.10\pm0.04$ \\
$\rm{[Fe/H],MW}$ & ($-$4.0, $-$0.2) & $-1.40\pm0.03$\\
 ${\rm{\sigma_{[Fe/H],satellite}}}$  & (0.001,10) & $0.48\pm0.03$ \\
 $\rm{\sigma_{[Fe/H],MW}}$ & (0.001,10) & $0.41\pm0.02$\\
$\mu_{\alpha}\cos \delta,_{\rm satellite} $ / 
  $\mu_{\rm{\delta,satellite}}$ (mas yr$^{-1}$) & ($-$3.5,3.0) / ($-$3.5,3.0) & $0.03\pm0.01$ / $-0.19\pm0.01$\\
$\mu_{\alpha}\cos \delta,_{\rm MW}$ / 
  $\mu_{\rm{\delta,MW}}$ (mas yr$^{-1}$) & ($-$3.5,3.0) / ($-$3.5,3.0) & $-1.41\pm0.05$ / $-0.33\pm0.06$\\
 $\sigma_{\mu_{\alpha} \cos \delta,_{\rm satellite}}$, $\rm{\sigma_{\mu_{\delta,satellite}}}$ (mas yr$^{-1}$) & fixed 0.025 \\
 $\sigma_{\mu_{\alpha} \cos \delta,_{\rm MW}}$,  $\rm{\sigma_{\mu_{\delta,MW}}}$ (mas yr$^{-1}$) & (0.1,19.95)/(0.1,19.95)& $0.72^{+0.04}_{-0.03}$ / $1.04\pm0.05$ \\
\enddata
\end{deluxetable*}

\begin{deluxetable*}{cccccccc}[!t]
\tablecaption{Properties for the Draco high probability ($p > 0.9$) member stars from GMM \label{tab:member}}
\tablewidth{0pt}
\tablehead{
\colhead{{\it Gaia} ID} &
\colhead{RA } &
\colhead{DEC } &
\colhead{$v\rm{_{hel}}$ } &
\colhead{$\rm{[Fe/H]}$} &
\colhead{$\mu_{\alpha} \cos \delta$} &
\colhead{$\mu_{\delta}$ }&
\colhead{Membership Prob} \\
\colhead{} &
\colhead{[Deg]} &
\colhead{[Deg]} &
\colhead{(km s$^{-1}$)} &
\colhead{} &
\colhead{(mas yr$^{-1}$)} &
\colhead{(mas yr$^{-1}$)}&
\colhead{}
}
\startdata
1433100054029203328 & 258.924736 & 57.618178 & $-295.76\pm1.23$ & $-2.51\pm0.10$ & $0.33\pm0.33$ & $-0.58\pm0.35$ & 1.0  \\
1433041470675928704 & 259.678035 & 57.533683 & $-287.77\pm2.71$ & $-2.48\pm0.18$ & $0.05\pm0.69$ & $-0.64\pm0.81$ & 0.998  \\
1433135478920752768 & 259.787367 & 57.568484 & $-303.70\pm2.04$ & $-2.02\pm0.10$ & $0.04\pm0.60$ & $0.54\pm0.66$ & 0.986  \\
1433131080874226048 & 259.834706 & 57.480566 & $-304.07\pm1.04$ & $-2.63\pm0.07$ & $-0.51\pm0.27$ & $-0.40\pm0.24$ & 0.998  \\
1433134796020830848 & 259.882808 & 57.559642 & $-304.68\pm0.53$ & $-3.20\pm0.03$ & $0.17\pm0.11$ & $-0.25\pm0.10$ & 1.0  \\
1432947084474633088 & 260.427404 & 57.549423 & $-293.47\pm1.60$ & $-2.12\pm0.13$ & $-0.01\pm0.39$ & $-0.69\pm0.39$ & 0.998  \\
1432931622592641152 & 260.762138 & 57.458133 & $-290.20\pm0.50$ & $-2.04\pm0.03$ & $0.09\pm0.14$ & $-0.30\pm0.12$ & 1.0  \\
1433115034875210880 & 259.055774 & 57.754213 & $-283.63\pm1.22$ & $-1.81\pm0.08$ & $-0.49\pm0.37$ & $-0.14\pm0.38$ & 0.974  \\
1433102184333023104 & 259.079197 & 57.689602 & $-275.55\pm3.68$ & $-2.77\pm0.50$ & $-0.38\pm0.68$ & $-0.77\pm0.67$ & 0.98  \\
1433109163656159232 & 259.231739 & 57.787975 & $-302.78\pm0.24$ & $-2.33\pm0.01$ & $-0.08\pm0.09$ & $-0.24\pm0.09$ & 1.0  \\
\enddata
\tablecomments{The entirety of this table is available online. \cwr{Additional data for this sample including distance to the center of Draco is available in \href{https://doi.org/10.5281/zenodo.17065707}{Zenodo}.}}
\end{deluxetable*}

\section{Results \label{results}}
The best fit parameters for the Draco heliocentric velocity, $\rm{[Fe/H]}$ and proper motion from the mixture model are listed in Table \ref{tab:prior}. \cwr{We also report uncertainties using the 16-84th percentile from the posterior distribution.}

\begin{figure*}
\centering
\includegraphics[width=15cm]{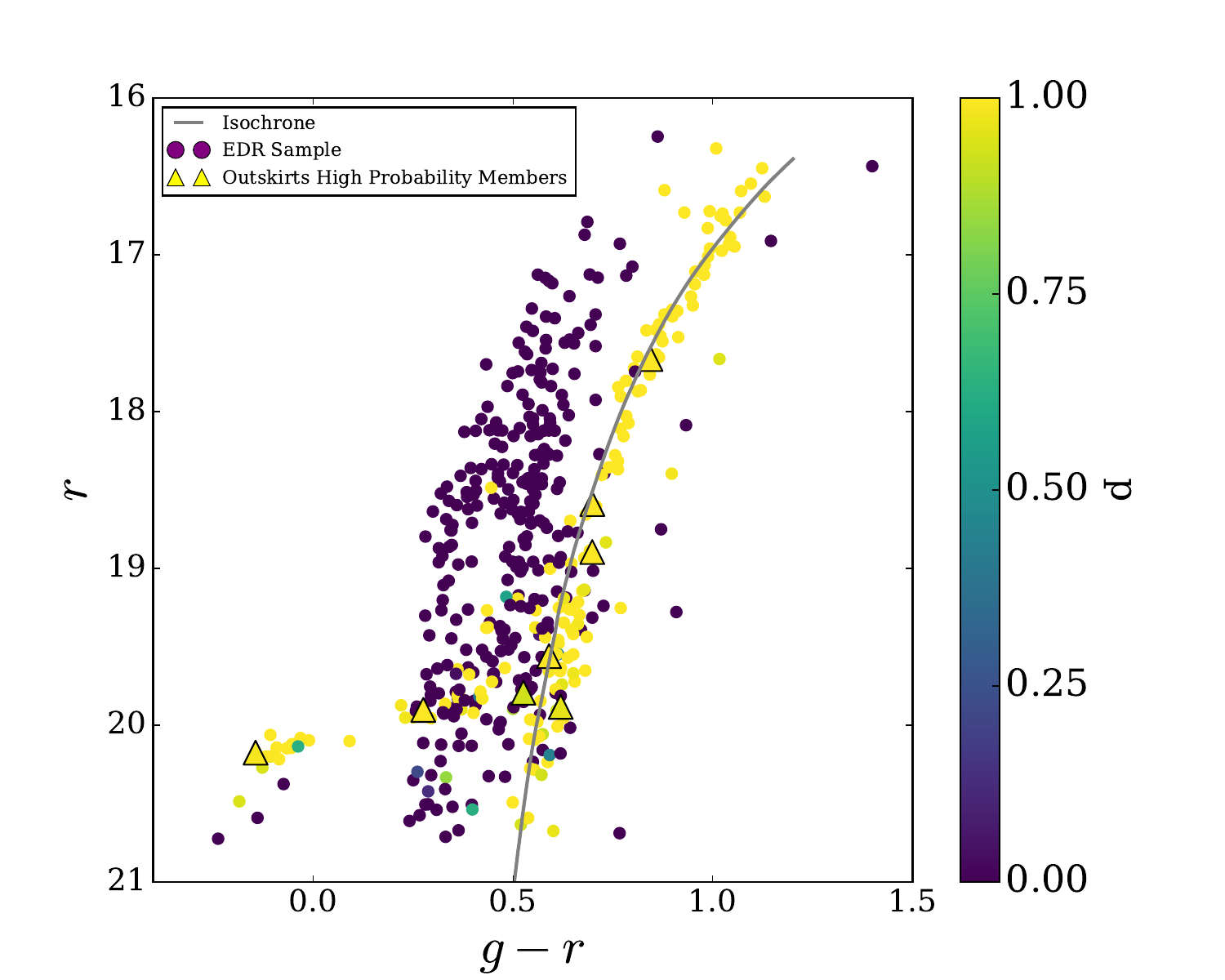}
\caption{The color magnitude diagram with the membership probabilities from the mixture model. Points are colored according
to their probability. The high probability members outside the tidal radius are labelled with triangles. \refco{The grey isochrone is for a metal poor population ([Fe/H] =-1.8, [$\alpha$/Fe] = 0.4) with age = 12.5 Gyr 
\citep{Dotter2008}.} \label{fig:prob}}
\end{figure*}

With the best fit mean and dispersion for $v\rm{_{hel}}$, $\rm{[Fe/H]}$, $\mu_{\alpha} \cos \delta$ and $\mu_{\delta}$ we use equation (7) to calculate the membership probability p$_{i}$ for the $i$-th star of the Draco candidates. 
The distribution of membership probabilities for the Draco candidates is shown in Fig. \ref{fig:probdis}.  We define high probability member stars to be those with probability $>$ 0.9, and the total number of high probability members is 155 stars. The list of all the high probability member stars is in Table \ref{tab:member}.
We plot the candidates, color-coded by their membership probabilities, in the color-magnitude diagram in Fig. \ref{fig:prob}.  The members outside the tidal radius are plotted as large triangles using the same color scale, \cwr{and are listed in Table \ref{tab:six}}. High membership probability stars are distributed along the isochrone, showing that stars identified as Draco members by the mixture model are also likely members based on their location in the CMD. 


\begin{deluxetable*}{ccccccc}[!t]
\tablecaption{Properties of Draco Members Outside the Tidal Radius Identified in this Analysis\label{tab:six}}
\tablewidth{0pt}
\tablehead{
\colhead{Star $\#$} &
\colhead{{\it Gaia} ID} &
\colhead{RA} &
\colhead{DEC} &
\colhead{GMM Probability} &
\colhead{Catalog\tablenotemark{a}} &
\colhead{Note\tablenotemark{b}} \\
\colhead{} &
\colhead{} &
\colhead{[Deg]} &
\colhead{[Deg]} &
\colhead{} &
\colhead{} &
\colhead{} 
}
\startdata
1 & 1432965054618434560 & 261.557746 & 57.725132  & 0.998 & DESI  \\
2 & 1434407579513539072 & 262.567080 & 57.721276 &  0.987 & DESI   \\
3 & 1433888300787161728 & 258.649105 & 58.117497 &  0.999   & DESI   \\
4 & 1433193345014967424 & 261.223205 & 58.291059 & 0.999 & DESI+Walker  \\
5 & 1434492516786370176 & 261.909752 & 58.263092 &  1.0  & DESI+Walker+Qi   \\
6 & 1433949770359264768 & 259.142131 & 58.498825 &  1.0  & DESI+Walker+Qi  \\
7 & 1433996705761927168 & 259.526634 & 58.513738 &  0.928  & DESI   \\
8 & 1433252164591733504 & 261.273875 & 58.628047 &  0.958  & DESI+Walker  \\
9 & 1434014607185724544 & 260.617774 & 58.730676 &  /  & DESI+Qi & $|v\rm{_{hel,satellite}}-\it v_{\rm hel}| > \rm 290 \ km \ s^{-1} $  \\
10 & 1420748587079854720 & 260.424144 & 56.060439  &  / & Walker & not in DESI SV area \\
11 & 1422045976440765568 & 262.811548 & 56.537112 &  /  & Walker & not in DESI SV area \\
12 &1433043051223687808 & 258.792829 & 57.271969 & / & Walker & no DESI SV fiber assigned \\ 
\enddata
\tablenotetext{a}{Walker stands for the \cite{Walker2023}  catalog analyzed in our work (see \S{\ref{5.2.1}}). Qi stands for the high probability extra-tidal stars identified in \cite{Qi2022}.}
\tablenotetext{b}{Reason why these members are not identified in our DESI sample.}
\end{deluxetable*}

\begin{figure}
\centering
\includegraphics[width=8cm]{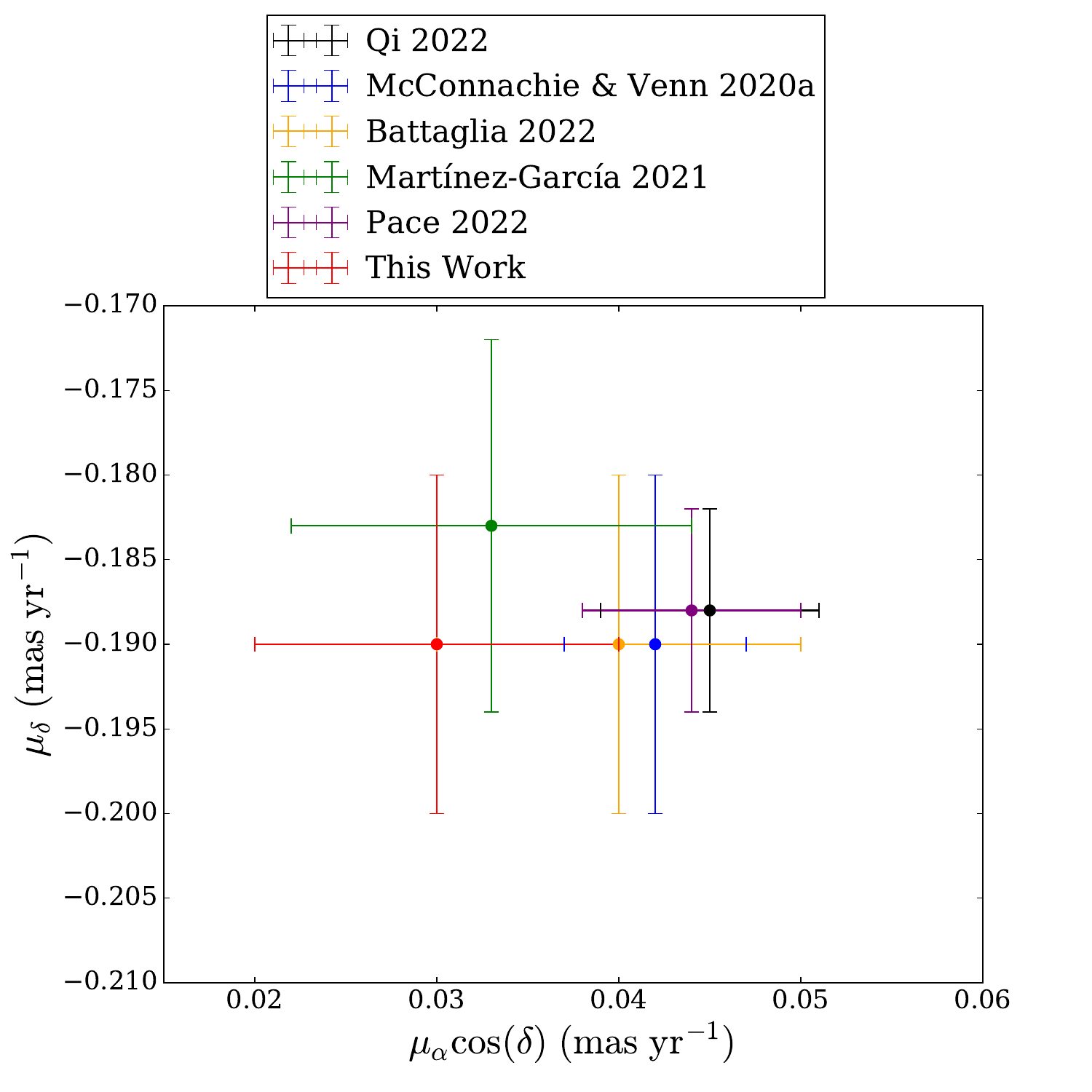}
\caption{Comparison of proper motion (PM) measurement between this work and previous literature. The red error bars show the Draco PM with 1-sigma uncertainties measured in this work. The black error bars are the PM value and uncertainty measured by \textcolor{black}{\cite{Qi2022}}. The blue error bars show the PM measured by \textcolor{blue}{\cite{McC2020}}. The orange data represents PMs from \textcolor{orange}{\cite{Batta2022}}, and the green crosses show PMs from \textcolor{green}{\cite{Mar2021}}. The purple PM measurement is from \textcolor{purple}{\cite{Pace2022}}. Our PM distributions are in consistent with other PM values.\label{fig:propernew}}
\end{figure}

\begin{figure*}
    \includegraphics[width=9.5cm]{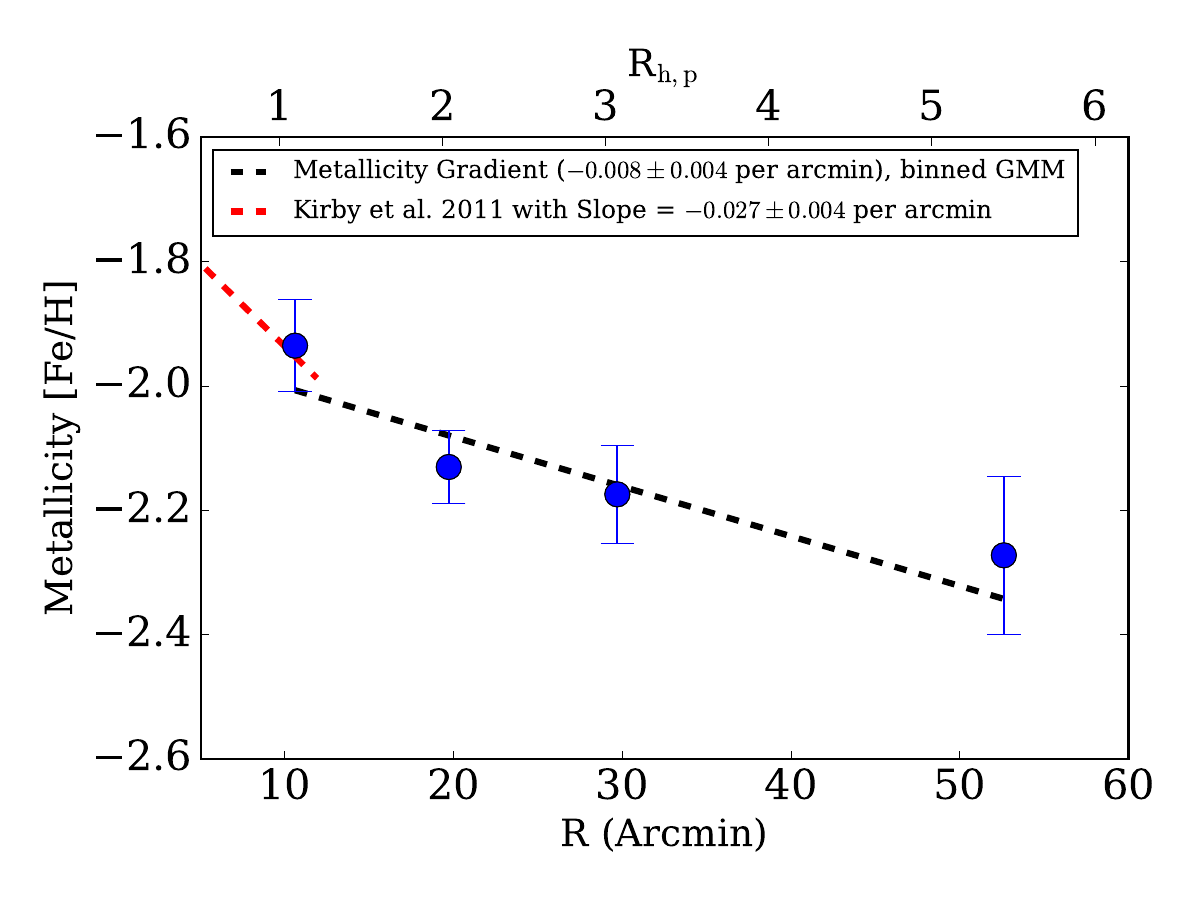}
    \hfill
    \includegraphics[width=9.5cm]{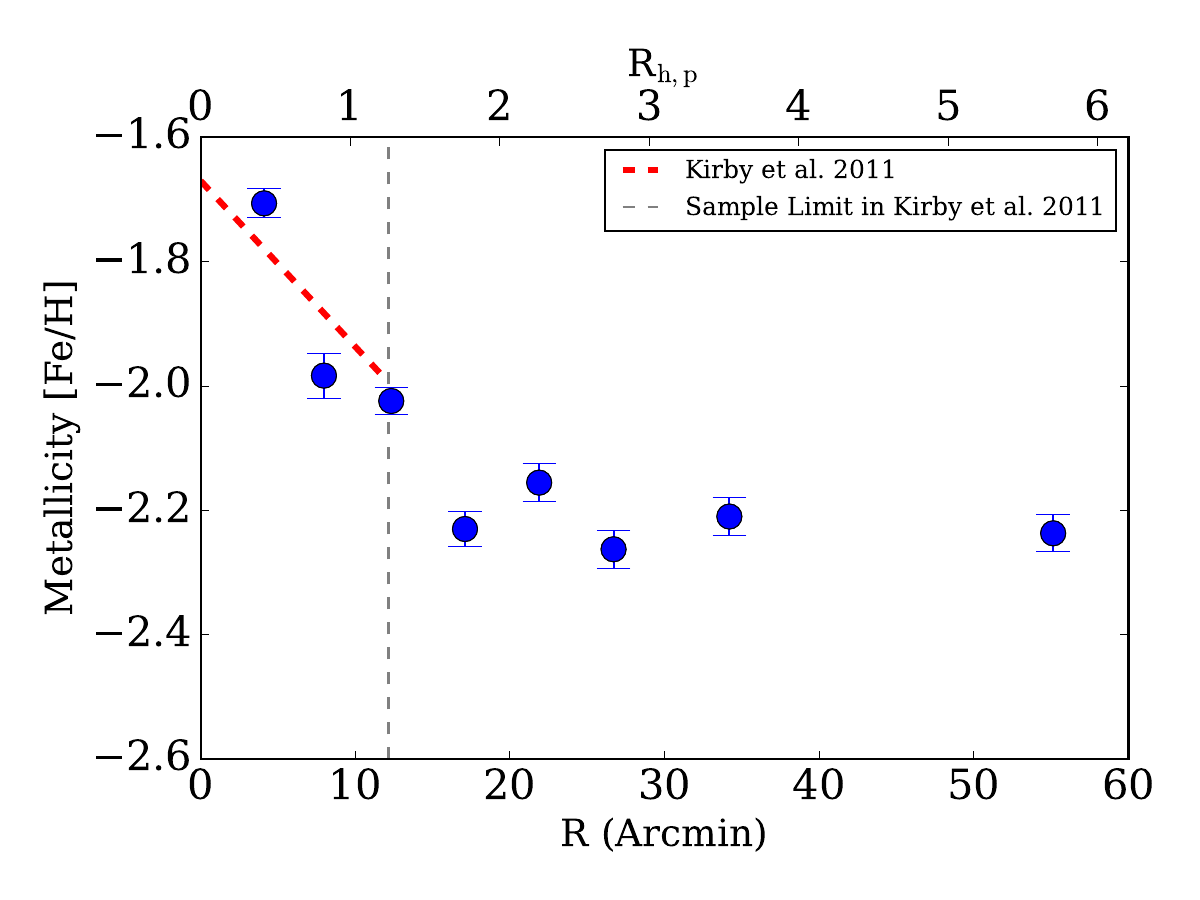}
        \caption{\refco{The metallicity gradient measured by two different methods of our high probability Draco members as a function of elliptical radial distance in units of half-light radius (top labels) and arcmin (bottom labels) from the center. {\bf{Left panel:}} The metallicity gradient measured by running the GMM model in four elliptical bins, fixing the proper motion. The best fit line of the radial distributions of [Fe/H] in this paper is labeled in dark dashed line. The slope of the metallicity distribution measured by \cite{Kirby2011} in the inner radius is labeled in red dashed line. {\bf{Right panel:}} The metallicity gradient as measured by computing the mean metallicity of just our high probability Draco members in elliptical bins as a function of the distance in units of half-light radius (top labels) and arcmin (bottom labels) from the center. The slope of the metallicity distribution measured by \cite{Kirby2011} in the inner radius is labeled in red. The slope of the metallicity measured by our high probability Draco members beyond 12 arcmin is $-$0.0006$\pm$0.0014 dex $\rm arcmin^{-1}$, consistent with being flat. The slope measured by \cite{Kirby2011} is a good fit within the half-light radius, and the metallicity gradient flattens at larger radii. }  \label{fig:metalgrad}}
\end{figure*}

\begin{figure}
\includegraphics[width=9.2cm]{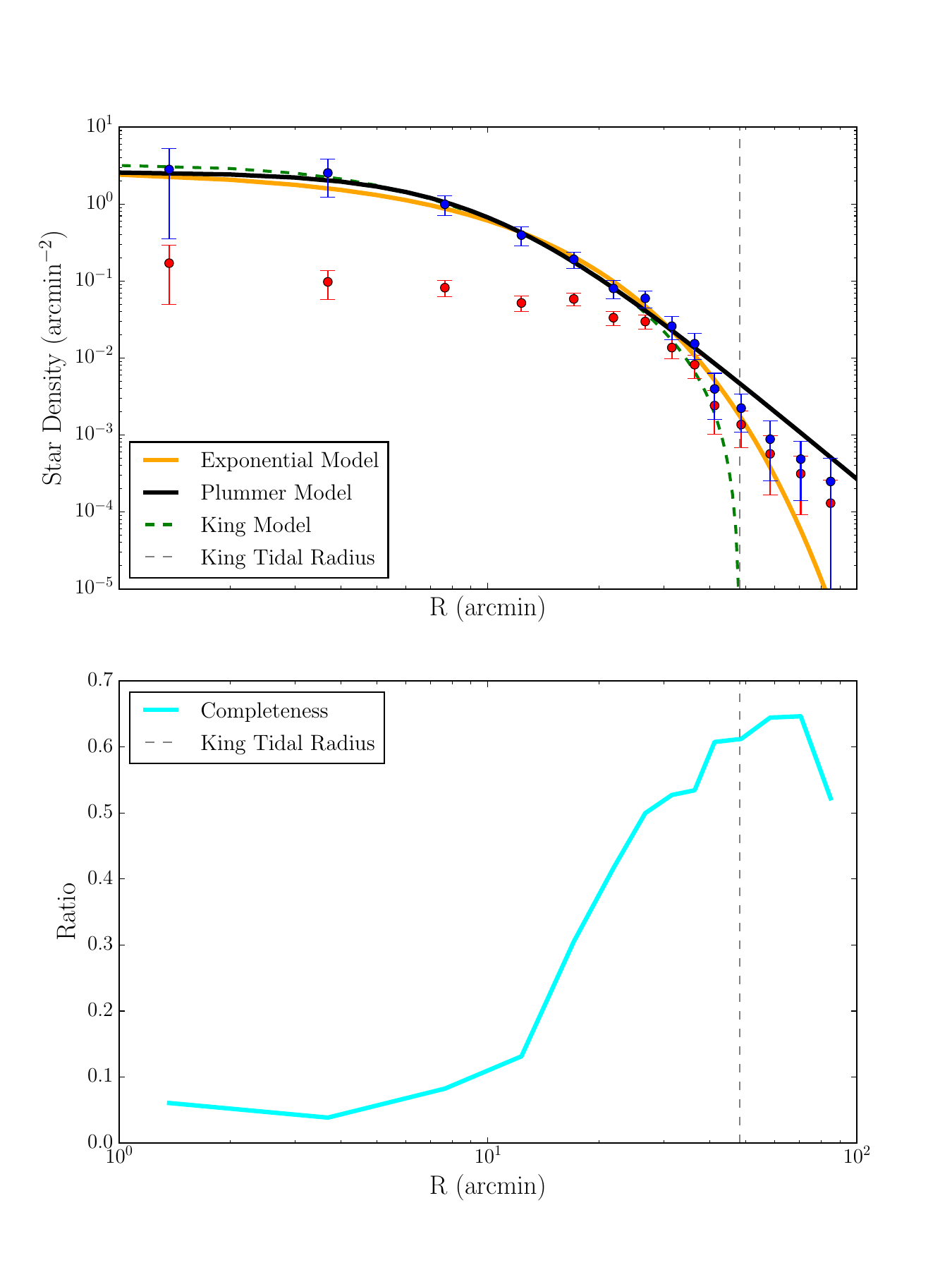}
\caption{{\bf{Top panel}}: The surface number density for stars with membership probability greater than 0.9, corrected for sample incompleteness (blue data points with error bars). A Plummer model is shown in black, a King surface density profile \citep{King1962} in green dashed and an exponential model in yellow solid. The red data points are the original measured surface density.  The grey dash line is the King tidal radius = 48.1 arcmin, as measured in \cite{Mu20182}. {\bf{Bottom panel}}: The completeness profile as a function of distance from center of Draco. The cyan line shows the ratio of the completeness. \label{fig:sd}
}\end{figure}

\begin{figure*}
\centering
\includegraphics[width=15cm]{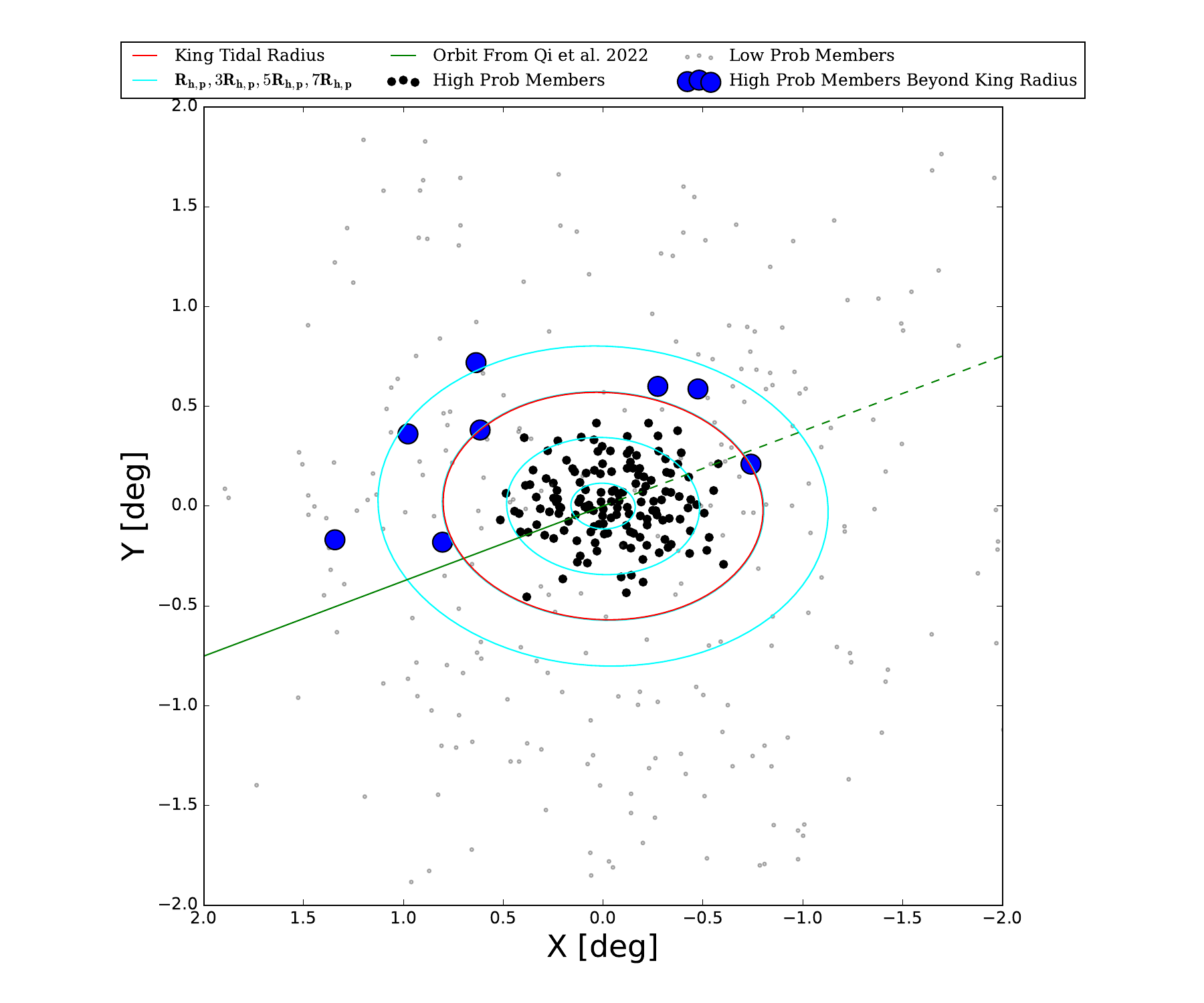}
\caption{Spatial distribution for the high probability Draco stars. The King tidal radius is the red ellipse. \refco{The orbit from \cite{Qi2022} is shown in green: the forward integration is plotted as a continuous line, while the backward integration is shown as a dashed line.} \cwr{This orbit was computed with a MW potential composed of a spherical Hernquist bulge \citep{Hernquist1990}, a spherical Hernquist nucleus, an axisymmetric Miyamoto–Nagai disc \citep{Miyamoto1975}, and a spherical Navarro–Frenk–White dark matter halo \citep{NFW1996}.}   The eight high probability stars outside the tidal radius are shown in larger circles labeled in blue. The high probability Draco members are labeled in dark circles while the low probability stars are labeled in grey circles. Ellipses at 1,3,5,7 times the half light radius \citep{Mu20182} are plotted in cyan. The ellipse at 5 times the half light radius nearly overlaps the King radius. \label{fig:spatial}}
\end{figure*}

\begin{figure}
    \includegraphics[width=9.cm]{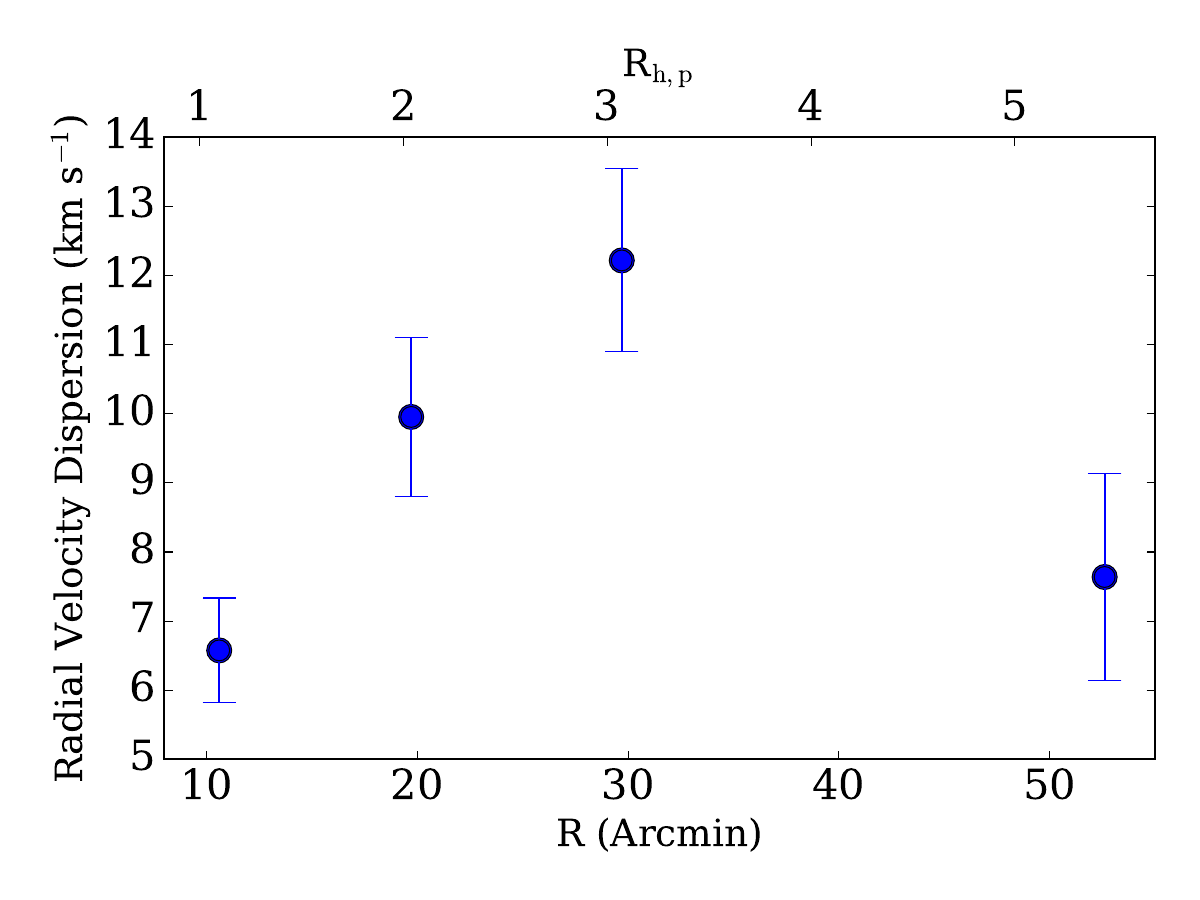}
    \caption{\refco{The velocity dispersion of our high probability Draco members as a function of elliptical radial distance in units of \comcorrect{half-light radius (top labels) and arcmin (bottom labels)} from the center. This was measured by running the GMM model in four elliptical bins, fixing the proper motion (same method as the metallicity gradient measurement in the left panel of Fig. \ref{fig:metalgrad}} ).\label{fig:vdispgrad}}
\end{figure}

\subsection{Heliocentric Velocity, Proper Motion and Metallicity Measurement}
\cwr{
Our mean line of sight heliocentric velocity for Draco is $v\rm{_{hel}}$ $= -290.62\pm0.80$ km s$^{-1}$ with a dispersion = $9.57^{+0.66}_{-0.62}$ km s$^{-1}$.  This mean velocity is consistent with the heliocentric velocity measured in \cite{Mu20182}, $v\rm{_{hel}}$ = $-291$ km s$^{-1}$. The Draco velocity dispersion measurement ($\rm{\sigma_{v_{hel}}}$ = $9.57^{+0.66}_{-0.62}$ km s$^{-1}$) is also consistent with the values quoted in previous work: $9.1\pm1.2$ km s$^{-1}$ in \cite{Mu20182}.}

Our measurements for the Draco dSph proper motion components are $\mu_{\alpha} \cos \delta$ = $0.03\pm0.01$ mas $\rm yr^{-1}$ and 
 $\mu_{\delta}$ = $-0.19\pm0.01$ mas $\rm yr^{-1}$.  Fig. \ref{fig:propernew} shows these values are consistent with previous studies of Draco that used only {\it Gaia} EDR3 proper motions and photometry. \cite{Qi2022} measured $\mu_{\alpha} \cos \delta$ = $0.045\pm0.006$ mas $\rm yr^{-1}$, $\mu_{\delta}$ = $-0.188\pm0.006$  mas $\rm yr^{-1}$ and \cite{Batta2022} found $\mu_{\alpha} \cos \delta$ = $0.04\pm0.01$ mas $\rm yr^{-1}$, $\mu_{\delta}$ = $-0.19\pm0.01$ mas $\rm yr^{-1}$.  




The best-fit value for the mean metallicity $\rm{[Fe/H]} =-2.10\pm0.04$ dex is slightly lower than the mean value found by \cite{Kirby2011} ($\rm{[Fe/H]} = -1.93\pm0.01$ dex) or \cite{Kirby2013} ($\rm{[Fe/H]} = -1.98\pm0.01$ dex). Our sample is more spatially extended, and that accounts for the difference in the best-fit [Fe/H].  As we show in Fig.~\ref{fig:metalgrad}, Draco is more metal poor at large radii, and stars at large radii make up most of our sample (see the completeness corrections in the bottom panel of Fig.~\ref{fig:sd}).  This leads to a lower mean metallicity measurement from our sample. This effect is also shown in \cite{Koposov2024}. We compute a weighted average [Fe/H] by using the completeness curve in Fig. \ref{fig:sd} to correct for the number of stars in each bin. The resulting completeness-weighted average [Fe/H] is $-1.91\pm0.02$ dex. \refco{The completeness-weighted average [Fe/H] computed within the same spatial extent used in \cite{Kirby2011} is $-1.83\pm0.05$ dex, consistent with their  measurement.}

Our measurement of the width of the metallicity $\rm{[Fe/H]}$ distribution  ($0.48\pm0.03$) agrees with previous measurements: $\rm \sigma_{[Fe/H]}$ = 0.47 in \cite{Kirby2011} and $\rm \sigma_{[Fe/H]}$ = 0.42 in \cite{Kirby2013}. 




\subsection{Spatial Distribution of the Sample}
\cwr{Our GMM does not use spatial information or a density profile constraint. This is done to minimize} the bias against identifying members near and beyond the tidal radius of the galaxy.
The spatial distribution of our sample with membership probability $>$ 0.9 is shown in Fig. \ref{fig:spatial} along with the orbit \citep{Qi2022} of the Draco dSph. The orthographic projection of RA and Dec to the Cartesian frame is:
\begin{align}
X = \cos \delta \sin(\alpha - \alpha_c),\\
Y = \sin \delta \cos \delta_{c} - \cos \delta \sin \delta_{c} \cos(\alpha - \alpha_c)
\end{align}

\noindent where $\alpha$ and $\delta$ are the location of a star in the RA and Declination directions, $\alpha_{c}$ and $\delta_{c}$ are the center coordinates of the dSph.
 The \cwr{red} ellipse is the Draco King tidal radius of 48.1 arcmin, as measured by \cite{Mu20182}.  The ellipse at the tidal radius is projected in the same way as discussed above, using the ellipticity value $\epsilon$ = 0.29 from \cite{Qi2022}. We detect eight stars outside this 48.1 arcmin tidal radius, extending as far as eight half light radii. The information for those members is listed in Table \ref{tab:six}.  
The spatial distribution for the high probability star members outside the tidal radius is shown in Fig. \ref{fig:spatial}.  These extra-tidal stars do not appear to be preferentially distributed along the orbit of Draco. Fig. \ref{fig:prob} shows that these members are distributed along an old, metal-poor, alpha-enhanced isochrone that matches the inner region of Draco ([Fe/H] = -1.8, age 12.5 Gyrs, [$\alpha$/Fe] = +0.4), further demonstrating their high probability of being members. Our RA/DEC selection region is 4$\times$5 degree$^2$, and the largest radius member is at 81 arcmin, \comcorrect{8.38 half-light radii}.

\subsection{Velocity Dispersion and Metallicity gradients}

We bin the sample relative to the center of Draco, from the center of the galaxy to 70 arcmin.  An elliptical radial binning with an adaptive bin size having at least 40 stars in each bin is applied to the sample. We then fit a GMM in each bin using the same likelihood in Equation (2) to measure the velocity dispersion and metallicity radial profiles of Draco, keeping \cwr{both the mean and dispersion of the proper motion components fixed.} \refco{The final velocity dispersion and metallicity as a function of distance from the center are shown in Fig. \ref{fig:vdispgrad} and the left panel in Fig. \ref{fig:metalgrad}.} 

\cwr{The radial velocity dispersion measurement does not show a steep decrease beyond a few half-light radii that would be expected if the total mass of the Draco dSph follows the light, consistent with Draco being embedded in a more massive dark matter halo.  Our velocity dispersion profile is consistent with the radial velocity dispersion profile we measure using the data from \cite{Walker2023} (see \S{\ref{s5.2.1}} for a description of our measurements using the \cite{Walker2023} data).}

\refco{In the left panel of Fig. \ref{fig:metalgrad}}, we show that our measurement of the metallicity gradient is shallower than the gradient measured by \cite{Kirby2011} of 0.027$\pm$0.004 dex per arcmin (red dashed line). The metallicity gradient in \cite{Kirby2011} is measured out to $r$ = 9 arcmin, near the half-light radius of Draco. Our sample of Draco members extends to much larger radii, and the metallicity profile does not appear to follow a smooth gradient.  To better compare our results to \cite{Kirby2011} inside and beyond the half-light radius, we divide our high probability members into eight elliptical radius bins and calculate the mean metallicity in each bin. \refco{The result is shown in right panel of Fig. \ref{fig:metalgrad}}. We measure the slope of the metallicity distribution for our high probability Draco members beyond 12 arcmin to be $-$0.0006$\pm$0.0014 dex $\rm arcmin^{-1}$, consistent with a flat slope. The metallicity gradient has a steeper slope within the half-light radius, more comparable to the value measured by \cite{Kirby2011}, and then becomes flatter at large radii.


\subsection{Surface Density Profile} 
We calculate the projected surface density per square arcmin of the high-probability Draco member stars using elliptical radial bins, with ellipticity $\epsilon$ =  0.29.  The DESI instrument has a fixed fiber density of 667/square degree, so this results in higher incompleteness in the central region of the dSph where the stellar density is the highest. \cwr{The high probability members are mainly from the SV survey, which makes up most of our spectroscopic sample.} To correct our density estimates, for each elliptical bin we calculate the ratio between the total number of targets available from the spectroscopic target selection described in \S{\ref{data}} and the number that were assigned to fibers and observed. \refco{We use that ratio to correct the surface density of our high-probability Draco members. The probability of a target being assigned to a fiber only depends on the local density of targets. Therefore, we assume that the mix of Draco and MW stars in our spectroscopic sample is the same as the mix in the full sample of targets at each elliptical radius.
} The 1D radial profile of our spectroscopic sample incompleteness is presented in the bottom panel of Fig.~\ref{fig:sd}. The corrected surface density is shown in blue in the top panel of Fig. \ref{fig:sd}, and the uncorrected surface density in red.

The corrected surface density is compared with the number density predicted from a King model profile, an exponential model and a Plummer model in the top panel of Fig. \ref{fig:sd}. 
\cwr{A King profile \citep{King1962} describes the density of a system in equilibrium in a tidal field, and assumes that mass follows light. 
\begin{equation}
\Sigma(R) = \Sigma_{0,\rm K} \left[ \frac{1}{\sqrt{1 + \left( \frac{R}{r_c} \right)^2}} - \frac{1}{\sqrt{1 + \left( \frac{r_t}{r_c} \right)^2}} \right]^2,
\end{equation}
where $r_t$ is the tidal radius, $r_c$ is the core radius, and $R$ is the elliptical radius as defined in \cite{Qi2022}. }
An exponential profile is usually a better fit to the MW satellite dSph surface density profiles \citep{MV2020}. 
\cwr{In this study, the exponential model is defined as:
\begin{equation}
   \Sigma\left(R\right)= \Sigma_{0,\rm E} \cdot e^{-\frac{R}{R_{0}}},
\end{equation}
where $\Sigma_{0,\rm E}$ and $R_{0}$ (exponential scale length) are the two parameters fitted to the data. }

We also compare the corrected density to the Plummer model \citep{Plummer1911}, which is:  
\begin{equation}
\Sigma\left(R\right)=\Sigma_{0,\rm P} (1+\frac{R^{2}}{r_{p}^{2}})^{-2},
\end{equation}
where $r_{p}$ is the Plummer scale length.
The exponential, King and Plummer model profiles are fitted to the corrected surface density measurements (blue in the top panel of Fig. \ref{fig:sd}) using the least squares method. 
The best fit exponential curve is labeled in yellow solid line, the King model is labeled in green dashed line and the Plummer model is leabled in dark solid line in Fig. \ref{fig:sd}. 

The uncorrected surface density measurements, plotted in red, are in approximate agreement with the models for large radii but very discrepant at smaller radii. The exponential model is a better fit to the Draco surface density near and outside the tidal radius. In \S{\ref{s5.1}}, we discuss the surface density and best fit model more in detail. 


\section{Discussion \label{discussion}}

\subsection{Surface Number Density at Large Radii} \label{s5.1}
The Draco surface number density profile is shown in the Top panel of Fig. \ref{fig:sd}.  Our high probability members extend to as far as eight half light radii from the center of Draco. Beyond 5 half light radii, the Draco surface number density profile shows a clear excess over the King profile and the single exponential profile.  We use our eight high probability Draco members beyond the King radius to measure the surface brightness of the Draco extra-tidal population in an elliptical annulus. The inner limit is at the King radius of 48.1 arcmin and outer limit is at 81 arcmin (8.38 half-light radius), the radius of our furthest detected Draco member. \refco{We use the luminosity function of an old metal-poor population isochrone with age = 12.5 Gyr, [Fe/H] = $-1.8$ \citep{Dotter2008} to estimate the 
total luminosity from the whole stellar population in the annulus, taking into account the incompleteness in our sample.} The surface brightness inside our elliptical annulus (corrected for sample incompleteness) is \refco{34.02} mag $\rm arcsec^{-2}$ in $V$ band. \cite{Shipp2023} quote a detectable surface brightness limit for stellar streams in the DES of 34 mag $\rm arcsec^{-2}$. We are able to identify this very low surface brightness extended structure of Draco because we can use [Fe/H], line of sight velocity and proper motion to select individual Draco member stars, rather than using only photometric data.


\cite{Sestito2023} discuss the surface density of stars outside the tidal radius for the Sculptor and Fornax dSph galaxies by using members provided in \cite{Jensen2024}. They find that Sculptor has an obvious excess above an exponential density profile that can be explained by a model of its tidal interaction with the MW. \cwr{Fornax is well fit by a single exponential function \citep{Sestito2023}, suggesting no recent tidal disturbance, and \cite{Jensen2024} do not find an extended stellar halo for Fornax.} We find that the surface density of Draco beyond its tidal radius has a small excess above an exponential profile, putting Draco between the cases of Sculptor and Fornax.  The orbital parameters of the three dSph galaxies also suggest the impact of tidal interaction with the MW on Draco is between that of Sculptor and Fornax.  From the orbital motion in \cite{Pace2022} taking into account the LMC, Draco has pericenter distance $r_\mathrm{peri}$ = 58 kpc and apocenter distance $r_\mathrm{apo}$ = 106.3 kpc. By comparison, Fornax has $r_\mathrm{peri}$ = 76.7 kpc and $r_\mathrm{apo}$ = 152.7.  Sculptor has $r_\mathrm{peri}$ = 44.9 kpc and $r_\mathrm{apo}$ = 145.7. \cite{Shipp2023,Shipp2024}
use the $FIRE$ and $Auriga$ simulations to show that disrupting satellites have pericenter $<$ 75 kpc and apocenter values $<$ 200 kpc,  consistent with Fornax showing the least tidal disturbance.  Sculptor passes closer to the center of the MW at apocenter than Draco, suggesting that Sculptor may have experienced a larger impact from tidal interactions with the MW potential than Draco \citep{Penarrubia2008,Gnedin1999}. 

Another possible explanation for the extended distribution of Draco stars we find is that it is the stellar halo of the Draco dSph. Simulations suggest dwarf galaxies form in extended dark matter halos and undergo mergers which create stellar halos \citep{Deason2022,Goater2024}. 
Low surface brightness, spatially extended structure has been discovered in several other dwarf galaxies \citep{Chiti2021,Yang2022,Jensen2024,Tau2024,Conroy2024}. The extended structure we identify in Draco is not aligned with its orbit, as would be expected if it originated from a tidal interaction with the MW. Instead, it may be a remnant stellar halo from a merger event earlier in Draco's formation. \comcorrect{\cite{Llambay2016} use simulations to show that dwarf-dwarf mergers can contribute to the formation of an outer metal poor stellar envelope through stellar accretion or dynamical heating. We find such an extended metal poor population in Draco (Fig. \ref{fig:metalgrad}), further evidence that it has had a merger history that can distribute stars to large radii and create the extended structure we see.}

\subsection{Comparison with Other Recent Draco Results \label{s5.2.1}} 
We compare our results with two other recent surveys of the Draco dSph by \cite{Walker2023} and \cite{Qi2022}. \cite{Qi2022} determine Draco membership probabilities using {\it Gaia} data. We descdribe the \cite{Walker2023} data we compare with below.
\subsubsection{Draco data in the \cite{Walker2023} spectroscopic catalog \label{5.2.1}}
\cite{Walker2023} present their latest reductions of the Magellan/M2FS and MMT/Hectochelle spectroscopic data on multiple dwarf galaxies, including Draco. In order to compare our outskirt members in Draco with those in the \cite{Walker2023} catalog, we apply our GMM analysis to their measured parameters $v_{\rm hel}$ (line of sight velocity) and {\it Gaia} DR3 proper motion.  
We use the same prior range as for the DESI data and we adopt the same cuts for the radial velocity and proper motion as discussed in Section \S{\ref{method}}.  In the right two panels of Fig. \ref{fig:fcom}, we compare [Fe/H], radial velocity and proper motion for the high probability members we find in our DESI plus {\it Gaia} sample with those we find in the \cite{Walker2023} catalog.  Using the data from \cite{Walker2023,Walker2008}, our best fit parameters are listed in Table \ref{tab:walker}. Our results from the DESI sample are in good agreement with the results from the sample of \cite{Walker2023,Walker2008}, allowing us to compare high probability members outside the tidal radius in the two samples.

\begin{deluxetable*}{cc}[!t]
\tablecaption{Values for the Best Fit Parameters for \cite{Walker2023} sample \label{tab:walker}}
\tablewidth{0pt}
\tablehead{
\colhead{Parameters} &
\colhead{Mean value with error}
}
\startdata
$v_{\rm{{hel}}}$ (km s$^{-1}$) & $-291.68\pm0.42$ \\
$\sigma_{v_{\rm hel}}$ (km s$^{-1}$) & $9.33^{+0.33}_{-0.32}$\\
$\mu_{\alpha} \cos \delta_{\text{walker}}$ (mas yr$^{-1}$) & $0.04\pm0.01$ \\
$\mu_{\delta,\text{walker}}$ (mas yr$^{-1}$)  & $-0.18\pm0.01$ \\
\enddata
\end{deluxetable*}

\subsubsection{Comparison of Draco Outskirt Members with Previous Studies}

\cwr{Table \ref{tab:six} lists all the stars identified outside the King tidal radius in our DESI sample as defined in \S{\ref{data}} or from our analysis of the \cite{Walker2023} catalog. We compare our DESI members with those from \citet{Qi2022} and the \cite{Walker2023} catalog below.}

\begin{figure*}
\centering
\includegraphics[width=15cm]{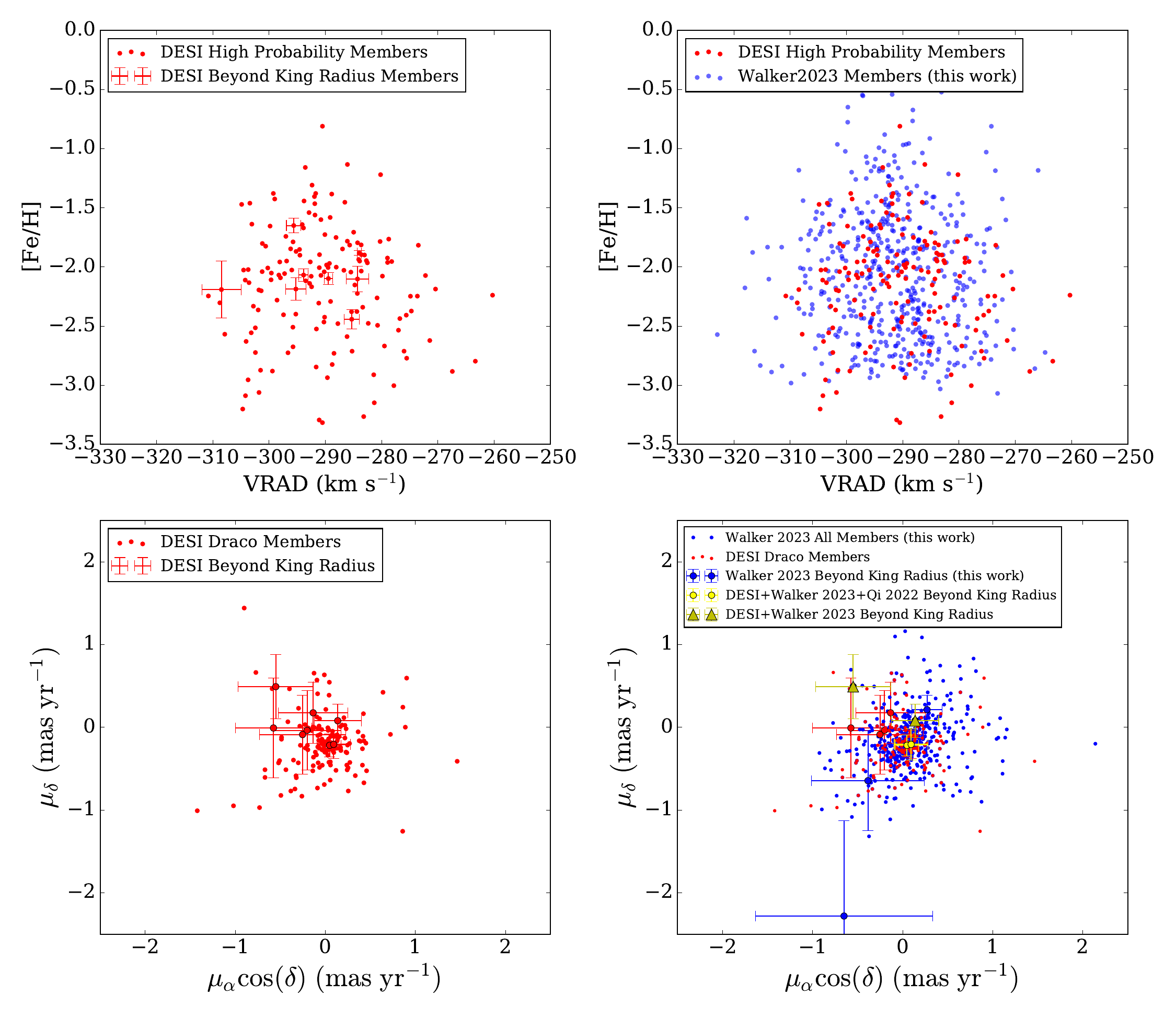}
\caption{[Fe/H] versus heliocentric radial velocity distribution and the proper motion distribution of high probability members in our work plotted with the high probability members from \cite{Qi2022} and high probability members we found using data from \cite{Walker2023}. {\bf{Top left panel}}: The [Fe/H] versus heliocentric velocity for the high probable members from our sample. {\bf{Top right panel}}: The [Fe/H] versus heliocentric velocity for the high probable members from our sample and \cite{Walker2023}. {\bf{Lower left panel}}: The proper motion distribution in this study. {\bf{Lower right panel}}: The proper motion distribution of our sample (in red circle) and high probability members in \cite{Walker2023} (blue circle). Two high probability members outside the tidal radius matched in all three studies (this work, \cite{Walker2023} and \cite{Qi2022}) are labeled in yellow circle. The yellow triangles show the two additional matched extra-tidal stars in our sample and in \cite{Walker2023} with the DESI errors for those two stars plotted in red cross\label{fig:fcom}.}
\end{figure*}

\cite{Qi2022} and our study have two members beyond the King tidal radius in common (star 5 and star 6 in Table \ref{tab:six}), shown in yellow circles in the lower right panel of Fig. \ref{fig:fcom}. These stars have metallicities and velocities measured by DESI and {\it Gaia} proper motions consistent with Draco and are distributed along an isochrone matching Draco's properties. Of the other seven members identified outside the tidal radius in \cite{Qi2022}, \cwr{for star 9 in table \ref{tab:six} with a {\it Gaia} ID of 1434014607185724544, we measure $v\rm{_{hel}}$ relative to the mean $v\rm{_{hel}}$ of Draco $>$ 290 km s$^{-1}$}, suggesting it is not a Draco member.  The others are outside the area of the DESI SV tile with the specially selected Draco SV sample, and they do not have spectroscopic observations in the DESI data we use for this study.

From our mixture model analysis of the \cite{Walker2023} data (see \S{\ref{5.2.1}}) we identify seven high probability Draco outskirt members. In the lower right panel of Fig. \ref{fig:fcom}, we plot the proper motion of the high probability members in our study and \cite{Walker2023}.  Four are in common with our DESI sample. Two of the four members are also identified by \cite{Qi2022} as discussed above. The other two are star 4 and star 8 in Table \ref{tab:six}. The $r$-band magnitude limit of our Draco spectroscopic data is about 0.5 mag fainter than \cite{Walker2023}.  We cross match the \cite{Walker2023} high probability outskirts members with our entire DESI Draco sample, regardless of membership probability.  We find that one of the extra-tidal stars we find in the \cite{Walker2023} sample (Star 12 in Table \ref{tab:six}) was in the area of the DESI Draco SV tile and passed the target selection criteria but was not assigned a fiber. The other two (Star 11 and 12 in Table \ref{tab:six}) are outside the area of the Draco SV sample and were not observed in the DESI sample used here. 

The members outside the tidal radius in our work, \cite{Qi2022} and \cite{Walker2023} are plotted in Fig. \ref{fig:fcom} with larger marker size. This panel also shows the proper motion distribution (labeled in yellow circles) of the two high probability members outside the tidal radius that overlap in the three studies: \cite{Walker2023}, \cite{Qi2022} and our sample.  The high-probability members outside the tidal radius from the DESI DR1 stellar catalog data and the \cite{Walker2023} data have a larger spread in proper motion compared to the members in \cite{Qi2022}. This may be because for both samples we have additional data, radial velocity and/or [Fe/H], to help determine the membership probabilities.  The lower left panel of Fig. \ref{fig:fcom} shows the proper motions and the proper motion errors of our DESI Draco members outside the tidal radius. They are still consistent with membership in Draco after taking into account the errors. 
The summary of the comparison of our extra-tidal Draco members with the other two studies are listed in Table \ref{tab:six}.

\section{Conclusion \label{conclusion}}
The DESI spectroscopic data from SV and the bright time survey covers 20 deg$^2$ centered on the dSph and extends to 21 in $r$-band. 
Utilizing heliocentric velocity and metallicity measurements from DESI over a wide area surrounding Draco, we are able to characterize the properties of the Draco dSph, with a focus on regions beyond its King tidal radius.

Combining line of sight velocity and [Fe/H] from DESI spectroscopy with proper motions from {\it Gaia} DR3,  we apply a GMM to the DESI data sample discussed in \S{\ref{data}}. The main results are as follows:
\begin{enumerate}
    \item \cwr{We identify 155 high probability Draco members in total. Among them, eight high-probability stars are outside the King tidal radius for Draco, two of which are in common with the previous studies of \citet{Qi2022} and \citet{Walker2023}.} 
    \item \cwr{The mean line of sight velocity is found to be $-290.62\pm0.80$ km s$^{-1}$ with a dispersion = $9.57^{+0.66}_{-0.62}$ km s$^{-1}$ and mean metallicity $\rm{[Fe/H]}$ = $-2.10\pm0.04$. These measurement are in good agreement with previous work.}
    \item \cwr{Draco exhibits a flat metallicity gradient beyond the half-light radius (\refco{Fig. \ref{fig:metalgrad}}). Within the half-light radius we measure value for Draco's metallicity gradient in good agreement with the previous results of \citet{Kirby2011} and \citet{Kirby2013}.} 
    \item  \cwr{The average surface brightness of stars outside the King tidal radius of Draco is \refco{34.02} mag $\rm arcsec^{-2}$. This is computed in an elliptical annulus extending from the tidal radius to 81 arcmin, the distance of the farthest member in our sample.} The orbital parameters of Draco suggest that it may not have experienced strong recent tidal interactions with the MW, and the extended structure in Draco is not aligned with its orbit. A possible explanation for the extended structure is that it is the stellar halo of Draco. 
\end{enumerate}

With data from DESI and other large-area spectroscopic surveys like 4MOST \citep{deJong2022}, WEAVE \citep{Jin2024} and PFS \citep{Tamura2016}, we can reproduce the experiment in this paper to revisit the extended stellar populations  in Draco. The methodology discussed in this paper can be applied to the future survey data of other MW dSph galaxies and shed light on a more comprehensive view of the extended stellar populations of these galaxies.

\section*{Acknowledgements}
This material is based upon work supported by the U.S. Department of Energy (DOE), Office of Science, Office of High-Energy Physics, under Contract No. DE–AC02–05CH11231, and by the National Energy Research Scientific Computing Center, a DOE Office of Science User Facility under the same contract. Additional support for DESI was provided by the U.S. National Science Foundation (NSF), Division of Astronomical Sciences under Contract No. AST-0950945 to the NSF’s National Optical-Infrared Astronomy Research Laboratory; the Science and Technology Facilities Council of the United Kingdom; the Gordon and Betty Moore Foundation; the Heising-Simons Foundation; the French Alternative Energies and Atomic Energy Commission (CEA); the National Council of Humanities, Science and Technology of Mexico (CONAHCYT); the Ministry of Science, Innovation and Universities of Spain (MICIU/AEI/10.13039/501100011033), and by the DESI Member Institutions: \url{https://www.desi.lbl.gov/collaborating-institutions}. Any opinions, findings, and conclusions or recommendations expressed in this material are those of the author(s) and do not necessarily reflect the views of the U. S. National Science Foundation, the U. S. Department of Energy, or any of the listed funding agencies.

The authors are honored to be permitted to conduct scientific research on Iolkam Du’ag (Kitt Peak), a mountain with particular significance to the Tohono O’odham Nation.

This work was supported by the U.S. Department of Energy, Office of Science, Office of High Energy Physics, under Award Number DE-SC0010107

APC acknowledges support from the Taiwanese Ministry of Education
Yushan Fellowship (MOE-113-YSFMS-0002-001-P2) and the Taiwanese
National Science and Technology Council (112-2112-M-007-009).

T.S.Li acknowledges financial support from Natural Sciences and Engineering Research Council of Canada (NSERC) through grant RGPIN-2022-04794.

SK acknowledges support from the Science $\&$ Technology Facilities Council (STFC) grant ST/Y001001/1.
For the purpose of open access, the author has applied a Creative Commons Attribution (CC BY) license to any Author Accepted Manuscript version arising from this submission.

\section{DATA AVAILABILITY}
The data and code used in this paper can be found in the \href{https://doi.org/10.5281/zenodo.17065707}{Zenodo repository}.

\bibliographystyle{aasjournal.bst}
\bibliography{MKY_2.bib}
\end{document}